\documentclass[pre,onecolumn,preprintnumbers,floatfix,amsmath,amssymb]{revtex4}
\usepackage{graphicx}
\usepackage{natbib}
\usepackage{subfigure}
\usepackage{color}
\usepackage{multirow}
\usepackage{amsmath}

\begin{document}

\title{Wetting and dewetting processes in the axial retraction of liquid filaments}
\author{Pablo D. Ravazzoli, Ingrith Cuellar, Alejandro G. Gonz\'alez, Javier A. Diez }
\affiliation{Instituto de F\'{\i}sica Arroyo Seco, Universidad Nacional del Centro de la Provincia de Buenos Aires, and CIFICEN-CONICET-CICPBA, Pinto 399, 7000, Tandil, Argentina}

\begin{abstract}
We study the hydrodynamic mechanisms involved in the motion of the contact line formed at the end region of a liquid filament laying on a planar and horizontal substrate. Since the flow develops under partially wetting conditions, the tip of the filament recedes and forms a bulged region (head), that subsequently develops a neck region behind it. Later on, the neck breaks up leading to a separated drop, while the rest of the filament restarts the sequence. One main feature of this flow is that the whole dynamics and final drop shapes are strongly influenced by the hysteresis of the contact angle typical in most of the liquid/substrate systems. The time evolution till breakup is studied experimentally and pictured in terms of a hybrid wettability theory which involves the Cox--Voinov hydrodynamic approach combined with the molecular kinetic theory developed by Blake. The parameters of this theory  are determined for our liquid/substrate  system (silicon oil / coated glass). The experimental results of the retracting filament are described in terms of a simple heuristic model, and also compared with numerical simulations of the full Navier--Stokes equations. This study is of special interest in the context of pulsed laser induce dewetting (PLiD).

\end{abstract}

\maketitle

\section{Introduction}
\label{sec:intro}
	
Controlling the size, shape and placement of nanoparticles is a main concern in many processes of nanostructures fabrication. For instance, the incorporation of plasmonic nanoparticles into photovoltaic devices can strongly increase their efficiency~\cite{atwater_nm10,wu_acsn11}, since the surface plasmon resonance between metallic nanostructures depends on the size and spacing~\cite{halas_cr11,le_acsn08} of the particles. Other applications include biodiagnostics and sensing, where Au nanoparticles bind to specific DNA markers, permitting their detection~\cite{rosi_cr05}. The potential applications for organized metallic nanostructures are wide ranging and include Raman spectroscopy~\cite{anker_nm08,vo-dinh_trac98}, catalysis~\cite{christopher_ar11}, photonics~\cite{ozbay_sc06}, and spintronics~\cite{wolf_sc01}. One strategy to create and organize structures at the nanoscale is to make use of the inherent self--assembly mechanisms of a material. The combination of natural instabilities related to the physical properties of liquid metals, such as low viscosity and high surface energy, with the technological ability to lithographically pattern nanoscale features, has created an open field of research. These techniques require the study of the liquid--state dewetting dynamics~\cite{dk_pof07} leading to liquid instabilities~\cite{wu_lang11}, with the goal of directing the assembly to precise, coordinated nanostructures in one~\cite{fowlkes_nanos12,fowlkes_nl14} and two~\cite{roberts_acsami13,fowlkes_nano11} dimensions. Within this context, the phenomena of wetting and dewetting of a shaped liquid film (metallic or not) covering a solid substrate are at the basis of all these processes and applications. However, even if these phenomena have been studied for the last two decades~\cite{ODB97,craster_09}, several of the main issues related to them are still under discussion. Therefore, one of the purposes of this paper is to provide a clear description of the wetting and dewetting mechanisms, including effects of contact angle hysteresis, and their implications in the formation of liquid structures.

\begin{figure}[htb]
\centering
\subfigure[]
{\includegraphics[height=2.5cm, width=5cm]{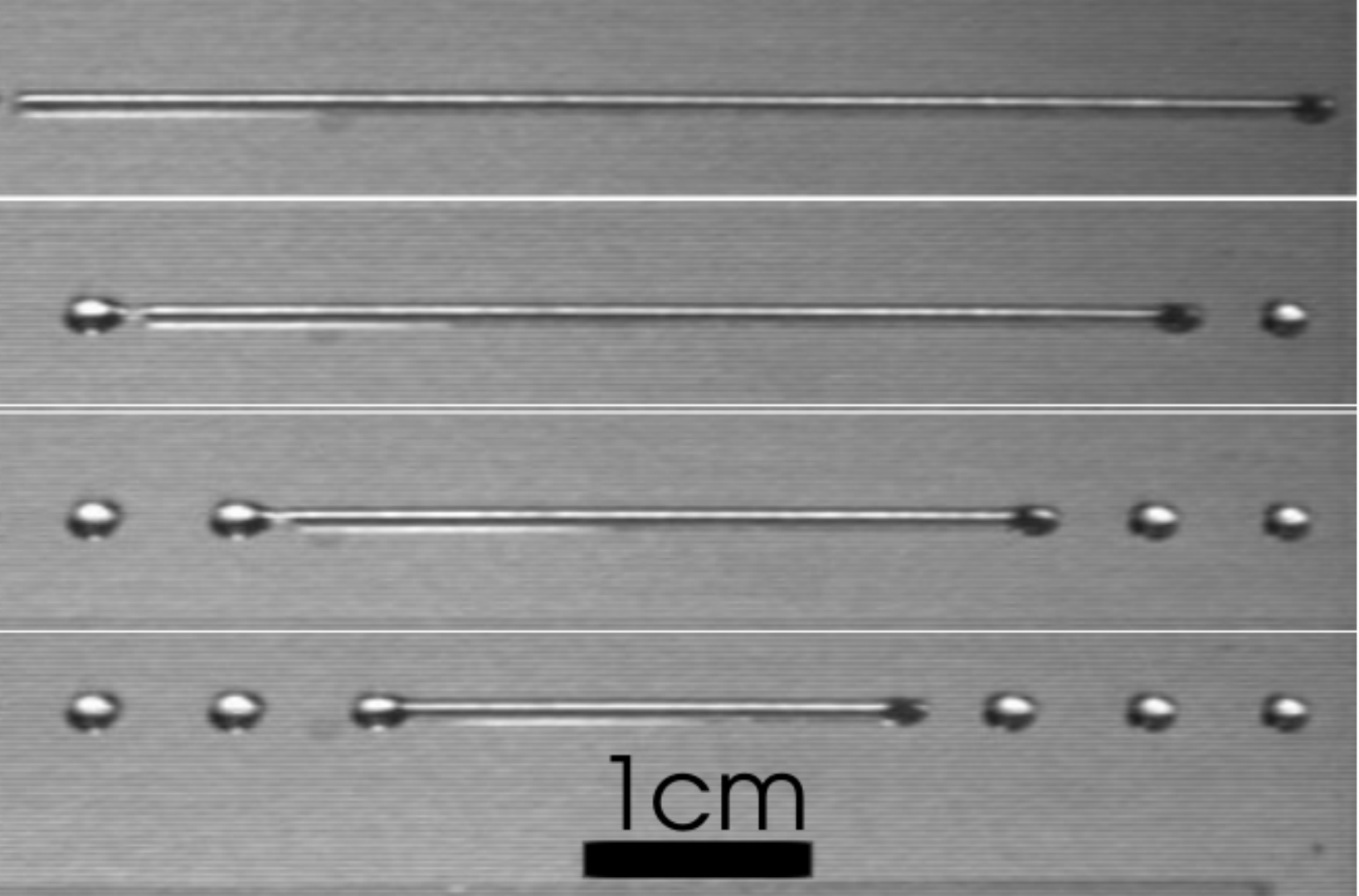}}
\subfigure[]
{\includegraphics[height=2.5cm, width=5cm]{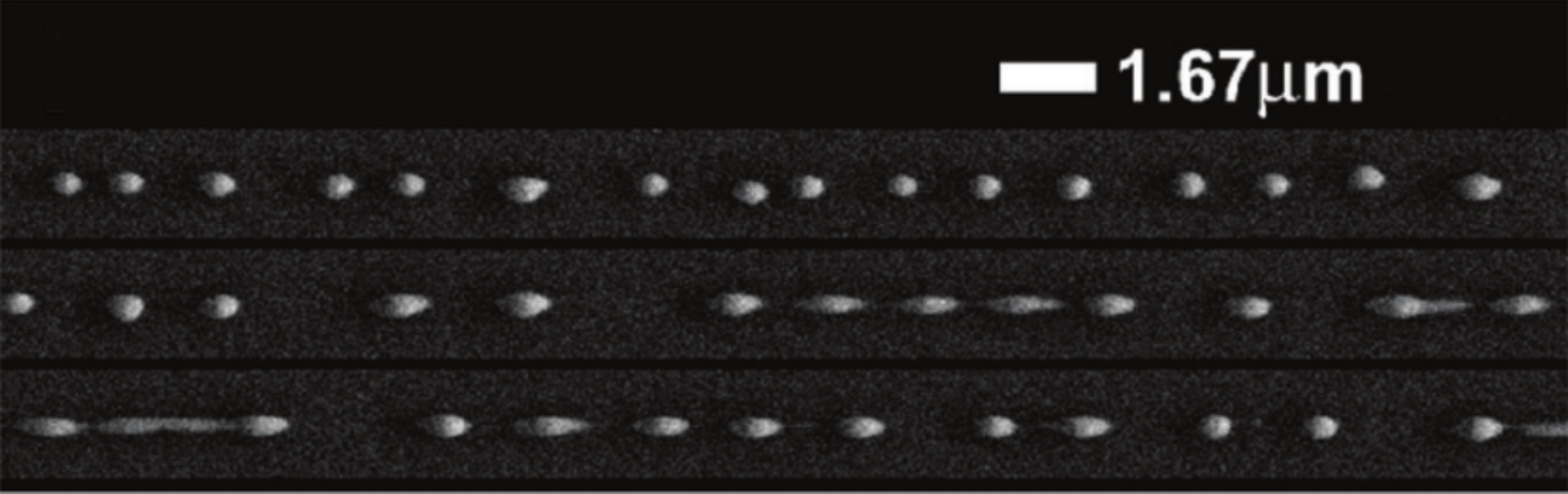}}
\caption{(a) Sequence of images at $t=0,240,602,958$~s showing the evolution of a liquid filament of width $w=0.123$~cm. (b) Similar process observed at nanoscale for a melted Ni filament~\cite{fowlkes_nano11}.}
\label{fig:sequence}
\end{figure}

We consider in detail the dynamics of a long liquid filament sitting on a horizontal substrate under partial wetting conditions. In particular, we focus our attention on the bulged region at its ends that is formed due to the filament retraction along its longitudinal direction. This process not only involves a dewetting motion, but also a wetting one in the transverse direction. Due to these advancing and receding motions at the ends, the contact line becomes curved in the region connecting the bulge (head) with the rest of the filament. This perturbation triggers an instability that leads to the appearance of a neck with rapidly decreasing width~\cite{gonzalez_07,dgk_pof12}. The breakup of this neck leads to two fluid portions which evolve separately. On one side, the head separated from the filament recedes until a final drop shape is achieved. The main features of this drop, such as its non--circular footprint, was thoroughly studied in previous works~\cite{rava_pof16, gonzalez_07}. On the other side, the remaining filament has a new shaped end as a result of the breakup process. The following receding longitudinal motion of this new end of the filament leads to another bulged region and a neck formation in a successive fashion. A group of pictures of this process, as obtained in this work for a silicon oil on a coated glass, is shown in Fig.~\ref{fig:sequence}(a). For comparison, we also include in Fig.~\ref{fig:sequence}(b) the analogous process observed at nanoscale~\cite{fowlkes_nano11,fowlkes_nanos12} for a filament of width $293$~nm and thickness $141$~nm, which was generated in the context of pulsed laser induce dewetting (PLiD) (see e.g.~\cite{fowlkes_nano11} and references cited there).

The main goal of this work is to describe and understand the physical mechanisms involved in the evolution of the head and its retraction. The flow before and after the breakup is studied experimentally by means of optical techniques, which allow to measure contact line positions, contact angles, thickness profiles and contact line contours. These data are then compared with a simple heuristic model as well as numerical simulations, which solve the Navier--Stokes equations with appropriate boundary conditions at the contact line. Note that the modeling of this flow implies the description of the contact line movement, and the subsequent pressure and velocity fields in a fluid domain confined by the substrate and the fluid free surface. Thus, the main issue in this modeling is how to determine the contact line motion, where both the stress and viscous dissipation rate diverge in the continuum description if slip is not allowed. Here, we consider a small amount of slip within the Navier boundary condition, and consistently add a relationship between the velocity and the contact angle. Since these relations must account for wetting and dewetting, we choose the combined approach proposed in~\cite{petrov_lang92}, which accounts for both the hydrodynamic of the macroscopic region~\cite{voinov_fd76} and the molecular kinetic~\cite{berg_93,blake_jcis69} for the contact line motion. The relevant physical parameters in this approach, which leads to a constitutive relationship for the contact line dynamics, are determined here by an additional study of the hysteretic behavior of a sessile drop.

In Section~\ref{sec:wet}, we obtain the main wettability properties of the substrate and liquid (silicon oil) system to be used in the filament experiments described in Section~\ref{sec:exp}. The experimental data are compared both with a simple heuristic model (Section~\ref{sec:model}) and numerical simulations of the Navier--Stokes equations (Section~\ref{sec:num}). Section~\ref{sec:conclu} is devoted to a summary and main conclusions of the work.

\section{Wettability of the system substrate/liquid}
\label{sec:wet}

In order to perform the experiments, we employ a similar setup to that reported in~\cite{rava_pof16}, i.e. we use a substrate that is partially wetted by our working fluid, namely a silicon oil (polydimethylsiloxane, PDMS). The substrate is a microscope slide (glass) which is coated with a fluorinated solution (EGC-1700 of 3M) using a Chemat Dip Coater under controlled speed. The PDMS partially wets the substrate, since the coating lowers the surface energy of the glass. In order to have reproducible wetting properties, we found that it is necessary to follow a protocol to get rid of the remaining solvent in the coating. Thus, the coated substrates are heated in an oven at $30^\circ$C for $30$ minutes, and left on a close and dry vessel for about $3$ days.

Both the surface tension, $\gamma$, and density, $\rho$, of the PDMS are measured with a Kr\"uss\ K11 tensiometer, while its viscosity, $\mu $, is determined with a Haake VT550 rotating viscometer. The values obtained for these parameters are: $\gamma = 21.0$~dyn/cm, $\rho = 0.97$~g/cm$^{3}$, and $\mu = 21.7$~Poise at temperature $T=23^\circ $C.

In order to characterize the wettability of the PDMS, we employ a technique which advances or recedes the  contact line by changing the volume of a sessile drop initially placed on the substrate~\cite{lam_acis02}. This is accomplished by injecting or withdrawing liquid through of a needle (connected to a syringe) in contact with the apex of the drop (see inset in Fig.~\ref{fig:hyst_static}(a)). The syringe volume is controlled by an Automated Dispensing System of a Rame--Hart Model 250 goniometer, and the measurements are based on the analysis of the axisymmetric drop shape profile (ADSA--P technique).

We distinguish here two scenarios: one in which the injection/withdrawal process is produced by pulses (between which the drop is allowed to reach a static shape), and another one where this process is practically continuous. Thus, in the first case we focus on the measurement of the static contact angle, $\theta_e$, while in the second case we are concerned with the dynamic angle, $\theta$, and its relationship with the drop contact line velocity.

\subsection{Static wettability}
\label{sec:st_wet}
We start with a drop of volume $V_0=25 \mu$l and use injection/withdrawal pulses of $1$~s duration at a flow rate $Q=\pm 1 \mu$l/s. This is done by means of an automated control of the piston motion in the syringe. The pulses are separated by time intervals of about $15$~s, so that the drop has time enough to relax to its static shape after every volume variation, $\Delta V=1 \mu$l. 

\begin{figure}[htb]
\centering
\subfigure[]
{\includegraphics[width=0.48\linewidth]{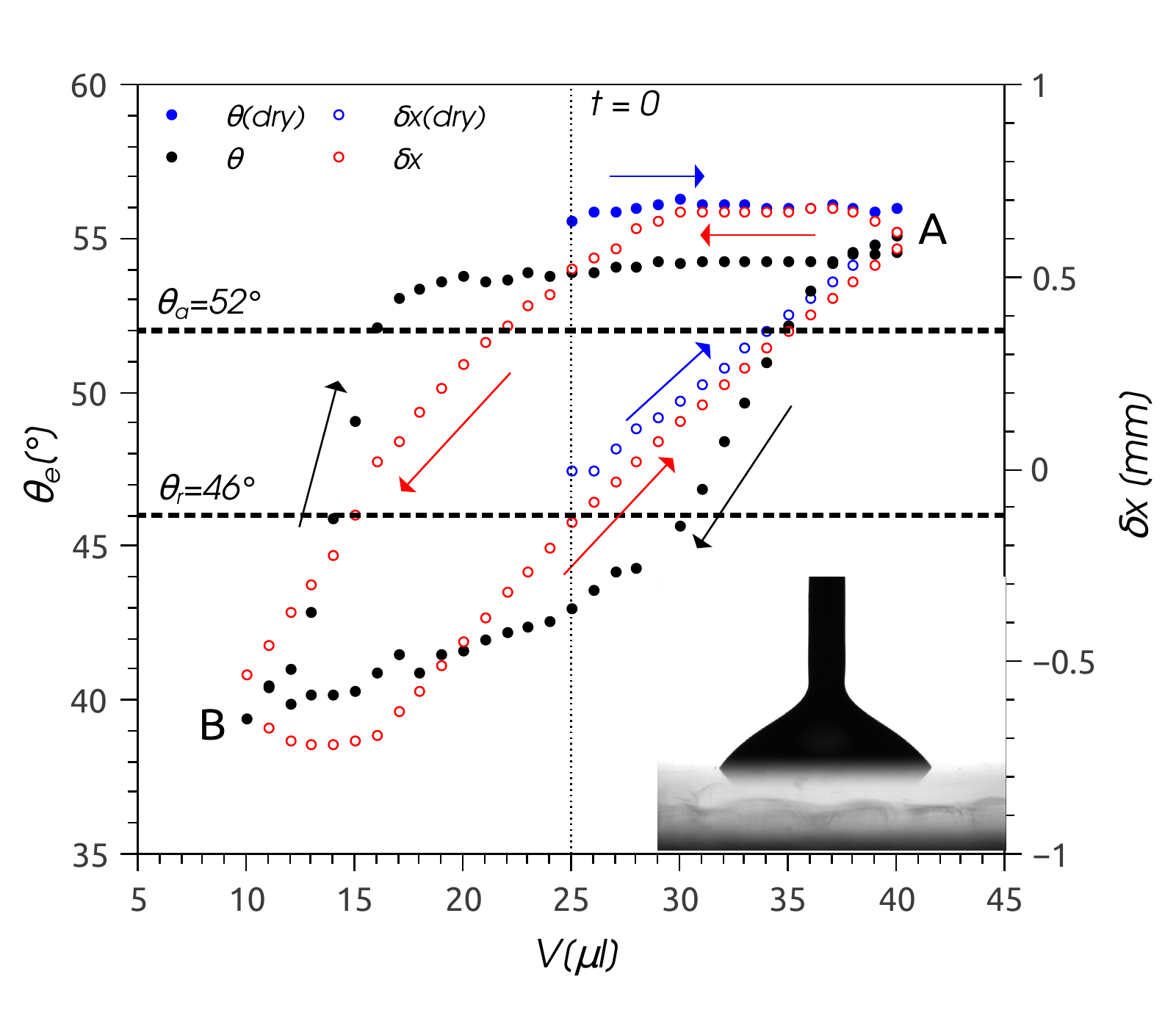}}
\subfigure[]
{\includegraphics[width=0.48\linewidth]{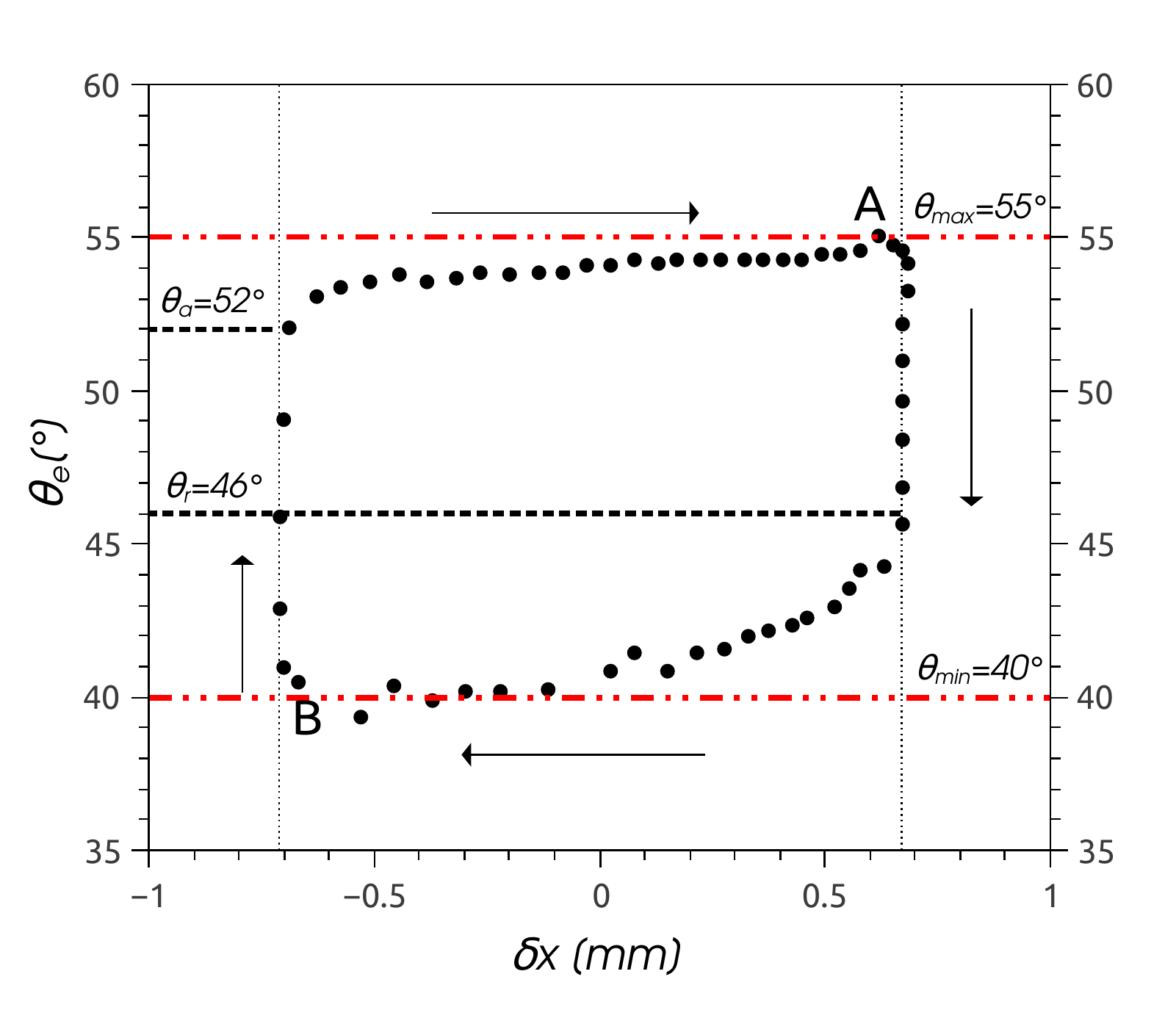}}
\caption{Hysteresis cycle of the static contact angle, $\theta_e$, (filled circles) and contact line displacement, $\delta x$, (hollow circles). (a) The drop initially spreads on a dry surface (blue circles) from $V_0=25$~$\mu$l till the maximum volume, $V_{max}=40$~$\mu$l, is reached. Then, $V$ varies from $V_{max}$ to $V_{min}=10$~$\mu$l and vice versa, while $\theta_e$ ($\delta x$) given by the filled black (hollow red) circles move in the clockwise (counterclockwise) direction, as the black (red) arrows indicate. The inset shows the drop profile on the substrate, and the needle at the drop apex . (b) $\theta_e$ versus $\delta x$. The volume varies clockwise from A-B (decreasing) and B-A (increasing), as indicated by the arrows.}
\label{fig:hyst_static}
\end{figure}

Initially at $t=0$, we have a sessile drop of volume $V_0$ with $\theta_e=56^\circ$ and null contact line displacement ($\delta x=0$). As the volume is increased, it spreads and reaches intermediate rest states on a dry surface (not previously covered by the PDMS) till the maximum volume, $V_{max}=40 \mu$l, is achieved. This is shown by the blue symbols in Fig.~\ref{fig:hyst_static}(a) for $\theta_e$ (full circles) and $\delta x$ (hollow circles), respectively. Note that $\theta_e$ remains practically the same for all these volumes, while $\delta x$ increases. This stage is not part of the hysteresis cycle, since only contact line motions on a previously wetted surface are considered in its determination. 

After this preliminary stage, the needle withdraws liquid until $V$ diminishes to $V_{min}=10$~$\mu$l (see the portion of the cycle from A to B in Fig.~\ref{fig:hyst_static}(a): clockwise for $\theta$, and counterclockwise for $\delta x$, as indicated by the arrows). Finally, the increasing volume stage (from B to A) starts at $V_{min}$ till $V_{max}$ is reached again. This process can be repeated indefinitely, but only the first hysteresis cycle is shown in Fig.~\ref{fig:hyst_static}(a) for brevity (the full black circles stand for $\theta_e$, while the hollow red ones correspond to $\delta x$). In order to fully understand the meaning of these curves in terms of the local behavior at the contact line, we plot $\theta_e$ versus $\delta x$ in Fig.~\ref{fig:hyst_static}(b), and ignore $V$ since it is not a relevant parameter for this analysis. Here, we see that the front starts to have a forward displacement (advance) at $\theta_a=52^\circ$ to achieve a new static position, so that this angle is called the (static) advancing contact angle. Similarly, we define the (static) receding contact angle,  $\theta_r=46^\circ$, as the value of $\theta_e$ at which the front must have a backwards displacement (recede) to achieve equilibrium. As it will be shown in the following sections, these two limiting angles are related to the dynamic contact angle behavior. Another relevant couple of angles is given by the maximum and minimum values of $\theta_e$, between which static drop states are possible. This range is $(\theta_{min},\theta_{max})=(40^\circ,55^\circ)$.

\subsection{Dynamic wettability}
\label{sec:dyn_wet}

\begin{figure}[hbt]
\centering
\subfigure[]
{\includegraphics[width=0.45\linewidth]{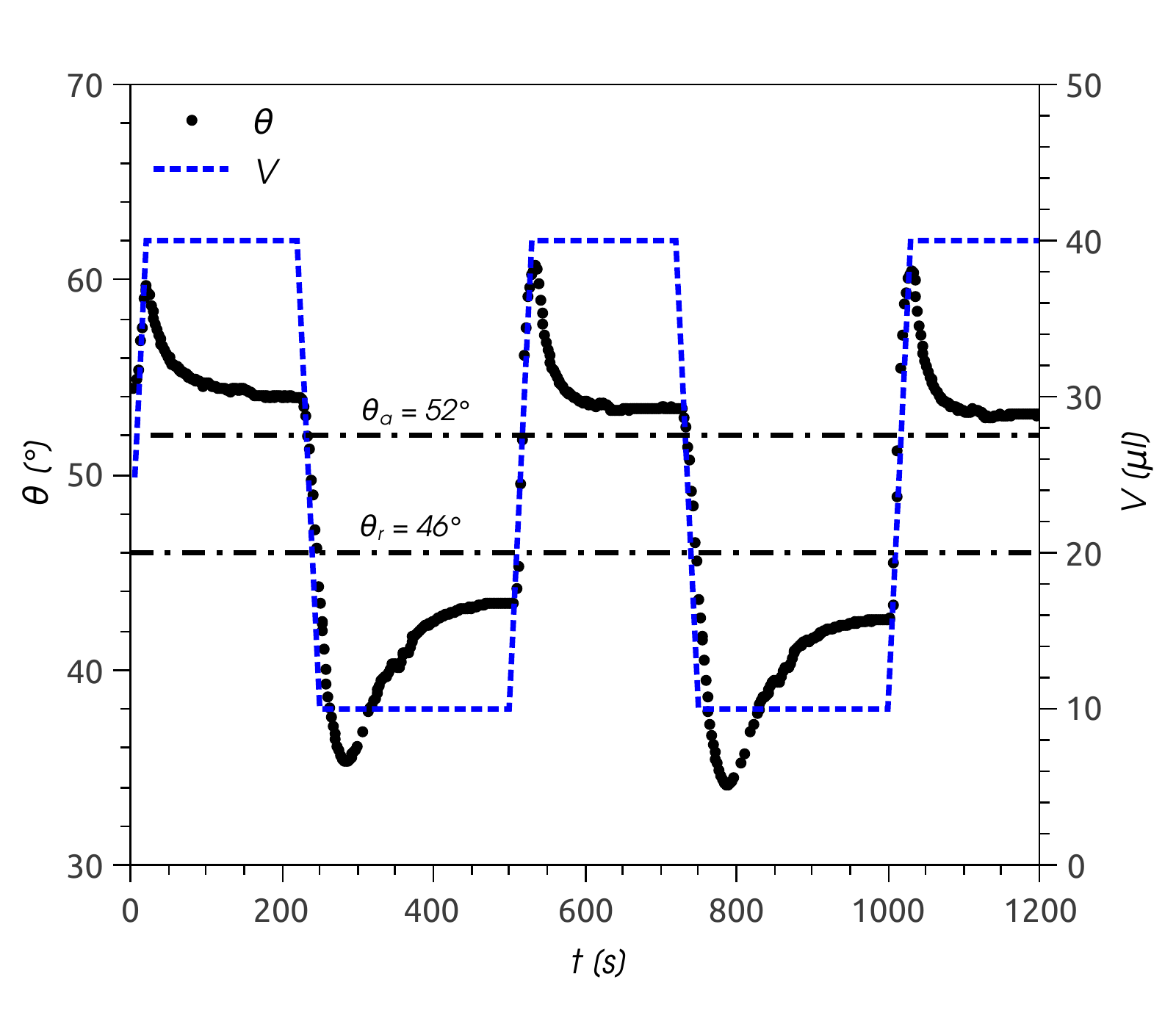}}
\subfigure[]
{\includegraphics[width=0.45\linewidth]{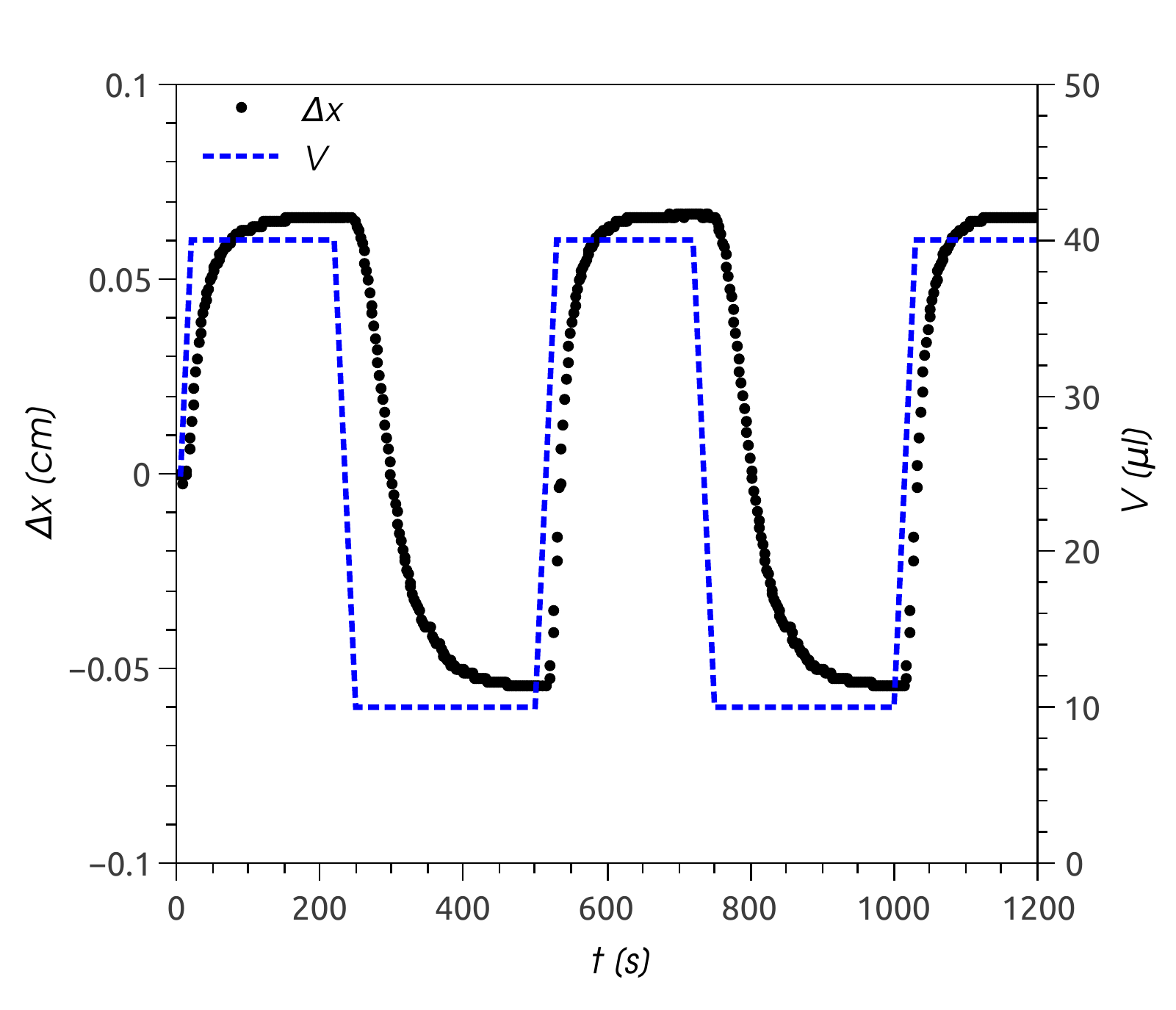}}
\caption{Time evolution of (a) the dynamic contact angle, $\theta$, and (b) the contact line displacement, $\Delta x$. The thin lines indicate the drop volumes for each stage, and the horizontal dotted lines in (a) correspond to the values of $\theta_a$ and $\theta_r$ obtained in the static regime.}
\label{fig:hyst_dyn}
\end{figure}

Now, we consider the relationship between the dynamic contact angle, $\theta$, and the contact line velocity, $v_{cl}$. Here, we inject/withdraw the liquid at the same flow rate, $Q$, as before, but with longer and larger pulses. Except for the first pulse (at $t=0$), which has a duration of $\Delta t=15$~s with a volume increase of $\Delta V=15$~$\mu$l, all the other ones last $\Delta t=30$~s with a volume variation of $\Delta V=\pm 30$~$\mu$l (see dashed lines in Fig.~\ref{fig:hyst_dyn}). Unlike the static case,  we continuously measure the dynamic contact angle, $\theta(t)$, and the corresponding contact line displacement,$\Delta x(t)$, as the drop spreads/contracts and relaxes to equilibrium for about $200$~s. The values of $\theta$ and $\Delta x$ as a function of time are plotted in Fig.~\ref{fig:hyst_dyn}. Note that during the injection and withdrawal stages $\theta$ varies in phase (without delay) with $V$, while the effect of $\Delta V$ on the contact line displacement, $\Delta x$, is delayed. This is due to the fact that during these stages the drop volume variation is quickly absorbed by a modification of the contact angle, while the contact line does not move till $\theta$ reaches the critical contact angle ($\theta_a$ for advancing, and $\theta_r$ for receding).

\begin{figure}[htb]
\centering
\includegraphics[width=0.6\linewidth]{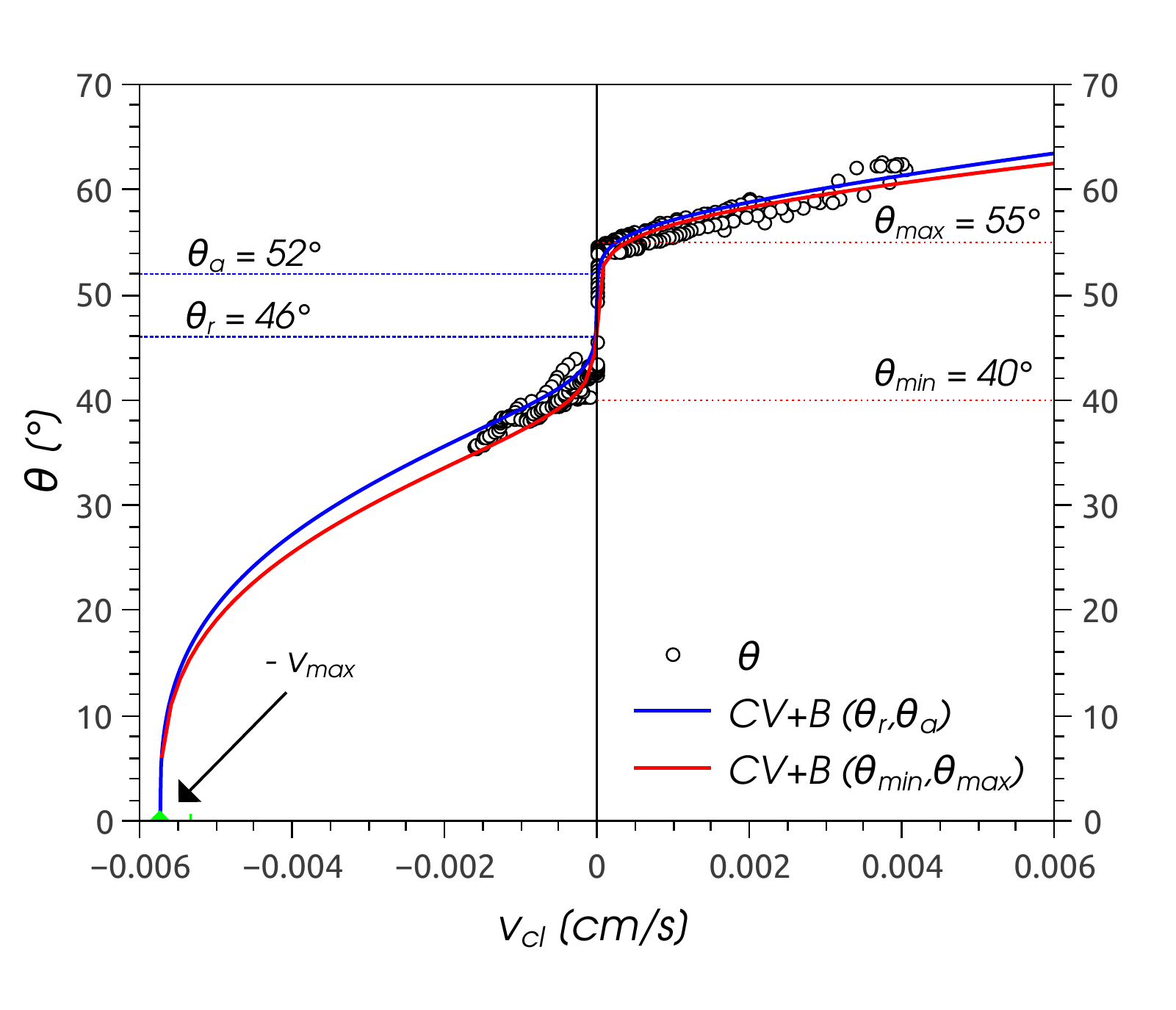}
\caption{Measured values of the dynamic contact angle, $\theta$, as a function of the contact line velocity of the drop, $v_{cl}$ (hollow circles). These data correspond to four cycles of the injection/withdrawal continuous flow at the drop apex. The arrow points to the maximum dewetting velocity, $v_{max}$, which corresponds to $\theta=0$.}
\label{fig:theta_vcl}
\end{figure}

We derive numerically the data of $\Delta x$ versus $t$ to obtain the contact line velocity, $v_{cl}(t)$. The hollow circles in Fig.~\ref{fig:theta_vcl} show $\theta$ as a function of $v_{cl}$, using $t$ as a parameter. Interestingly, the $\theta$ range for which $v_{cl}=0$ is practically coincident with the hysteresis range of $\theta_e$, namely $(\theta_r,\theta_a)=(46^\circ,52^\circ)$ shown in Fig.~\ref{fig:hyst_static}(b). This agreement confirms that the methodology used to determine both types of angles, namely static and dynamic, is appropriate to describe the wettability of the substrate by the PDMS.

In order to fully describe the dewetting stage, we must enlarge the velocity range for $v_{cl}<0$. In fact, it is known that the dewetting velocity has a maximum absolute value at which the dynamic contact angle is zero~\cite{eral_13,petrov_lang92}, and our results in Fig.~\ref{fig:theta_vcl} for the dynamic cycle of the sessile drop cannot show this phenomenon. 

With this goal, we perform a different experiment which allows to obtain the maximum velocity of dewetting, referred here to as $v_{max}$ ($>0$). In this new setup, we plunge a coated substrate into a deep pool of PDMS, and withdraw it at a speed, $V_s$, greater than $v_{max}$. This is achieved when, for a certain $V_s$, the liquid is entrained off the pool, and forms a film on the substrate. Once the substrate has stopped, the contact line of this film starts to slide down the surface due to gravity. This dewetting motion proceeds at the maximum possible velocity, $v_{max}$, with zero contact angle~\cite{petrov_cs85}. We perform experiments with different withdrawal speeds ($0.033<V_s<0.5$~cm/s), and confirmed that the value of $v_{max}$ is independent of $V_s$, as expected. The contact lines on both sides of the coated glass are seen on the digital image due to the transparency of the substrate. They are not exactly coincident, thus introducing some error in the measurement, which is at most $7\%$ and increases as the front dewets. Finally, by measuring the contact line position as a function of time, we are able to obtain the linear regression expression 
\begin{equation}
 v_{max}= (5.73 \pm 0.24)\times 10^{-3}\, cm/s.
 \label{eq:vmax}
\end{equation}
This value of $v_{max}$ is shown in Fig.~\ref{fig:theta_vcl} for $\theta=0$, and it is of great importance to determine the parameters of the physical mechanism that describes the $\theta$--$v_{cl}$ relationship, as will be shown in the following section.

\subsection{Wettability model}
\label{sec:wet_model}
In order to describe the motion of the contact line over the substrate ($xy$-plane), we employ two constitutive relations to account for the liquid--solid interaction. Firstly, we overcome the stress singularity by relaxing the no slip boundary condition at the substrate through the Navier formulation (see e.g.~\cite{Haley91}),
\begin{equation}
 v_{x,y} = \ell \, \frac{\partial v_{x,y}}{\partial z}\quad \text{at $z=0$,} 
 \label{eq:slip}
\end{equation}
where $\ell$ is the slip length. This parameter embraces somehow the main features related with the intermolecular forces in that region. Secondly, we use a relationship between the dynamic contact angle, $\theta$, and the contact line velocity, $v_{cl}$, which is normal to the contact line itself. Since we need to describe not only advancing motions (wetting,  $v_{cl}>0$), but also receding (dewetting,  $v_{cl}<0$) ones, we define the dynamic contact angle from a combination of the hydrodynamic and molecular kinetics models~\cite{berg_93,blake_jcis69,blake_jci06} as first proposed in~\cite{petrov_lang92}. For the former, we have the Cox-Voinov relationship~\cite{voinov_fd76,cox_jfm86}
\begin{equation}
 \theta^3 = \theta_m^3 + 9 \frac{\mu v_{cl}}{\gamma} \ln \left( \frac{L}{\ell} \right)
 \label{eq:theta_hyd}
\end{equation}
where $\theta_m$ is the microscopic contact angle, and $L$, $\ell$ are macroscopic and microscopic length scales, respectively. 
Usually, the former is assumed to be a constant given by Young's law, and independent of $v_{cl}$. Here, we take $L$ as the capillary length, i.e.
\begin{equation}
 L=a_c=\sqrt{\frac{\gamma}{\rho g}}
\end{equation}
where $g$ is the gravity. Thus, we leave $\ell$ as a parameter to be determined by adjusting the experimental data from the final modeling.

Note, however, that Eq.~(\ref{eq:theta_hyd}) with $\theta_m=const.$ is not appropriate to adjust well the experimental data in Fig.~\ref{fig:theta_vcl}, which show a sharp jump at $v_{cl}=0$ due to hysteresis effects. Then, we must resort to an alternative approach. In fact, several authors~\cite{cox_jfm86,voinov_fd76} admit the possibility of a non--hydrodynamic velocity dependence of $\theta_m$. Here, we consider that it is given by the Blake molecular--kinetics model~\cite{blake_jcis69} in the form,
\begin{equation}
 \cos \theta_m = \cos \theta_0 - \frac{1}{\Gamma} \sinh^{-1} \left( \frac{v_{cl}}{v_0} \right)
\label{eq:theta_bl}
\end{equation}
where $\theta_0$ is the (microscopic) equilibrium contact angle ($v_{cl}(\theta_0)=0$). The other parameters, which are of molecular origin, are given by~\cite{blake_jcis69}
\begin{equation}
 v_0= 2 \kappa \lambda, \qquad \Gamma= \frac {\gamma \lambda^2}{2 k T}
 \label{eq:v0_Gamma}
\end{equation}
where $\kappa$ is the frequency of molecular displacement at equilibrium, $\lambda$ is the average length of each molecular displacement (or distance between adsortion sites), $T$ is the temperature, and $k$ is the Boltzmann's constant. Since we do not focus our attention in the kinetic process itself, we consider $v_0$ and $\Gamma$ as fitting parameters of our modeling. On the other hand, since the surface energies are usually associated with the cosine of a contact angle, as in Young's law, we consider the cosine of the equilibrium angle, $\theta_0$, as an average of the cosines of two characteristic angles of the hysteresis range. If we take the  static advancing and static receding contact angles, $\theta_a$ and $\theta_r$, respectively, we have
\begin{equation}
 \cos \theta_0 = \frac{\cos \theta_{a}+\cos \theta_{r}}{2},
 \label{eq:theta0}
\end{equation}
which yields $\theta_0=49.07^\circ$. We could also have taken the pair $(\theta_{min},\theta_{max})$ instead of $(\theta_r,\theta_a)$, in which case the angle would be $\theta_0^\ast =47.95^\circ$. However, as we will discuss later, this choice does not lead to any improvement in the comparison with the experimental results.

Note that both Eqs.~(\ref{eq:theta_hyd}) and (\ref{eq:theta_bl}) account for the hysteresis of the contact line, since they provide the dynamic contact angle, $\theta$, as a function of $v_{cl}$.  In order to determine the parameters $v_0$, $\Gamma$, and $\ell$, we fit the data from the dynamic cycling contact angle measurements (circles in Fig.~\ref{fig:theta_vcl}), and the value $(v_{max},0)$ (rhombic symbol  in Fig.~\ref{fig:theta_vcl}, and Eq.~(\ref{eq:vmax})) with the approximating function 
\begin{equation}
 \theta^3 = \arccos^3 \left[ \cos \theta_0 - \frac{1}{\Gamma} \sinh^{-1} \left( \frac{v_{cl}}{v_0} \right) \right] + 
 9 \frac{\mu v_{cl}}{\gamma} \ln \left( \frac{a_c}{\ell} \right).
 \label{eq:theta_fit}
\end{equation}
This combined Cox-Voinov-Blake model (for brevity {\it CV+B}, or hybrid model) is solved with the following iterative procedure. For a given pair $(\Gamma^{(0)},\ell^{(0)})$ we obtain $v_0^{(0)}$ such that $\theta (v_{max})=0$; then, with this value of $v_0^{(0)}$ we calculate a new pair $(\Gamma^{(1)},\ell^{(1)})$ as given by a square minimum method using Eq.~(\ref{eq:theta_fit}) with the dynamic cycling measurements. By using this pair, we determine a new value of $v_0^{(1)}$ such as $\theta (v_{max})=0$, and so on. After some tens of iterations, we obtain the converged values
\begin{equation}
 \Gamma= 69.4455, \quad \ell= 0.0026535 a_c=3.94 \times 10^{-4}\, cm , \quad v_0=2.4511 \times 10^{-6}\, cm/s.
 \label{eq:param}
\end{equation}
The best fitting curve given by Eq.~(\ref{eq:theta_fit}) with these parameters is shown in Fig.~\ref{fig:theta_vcl} (see blue line). The red line is the corresponding curve using $\theta_0^\ast$ instead of $\theta_0$. Even if the  difference between them is small and is mainly noticeable in the dewetting region, we will use in what follows the parameters in Eq.~(\ref{eq:param}) for the blue curve in Fig.~\ref{fig:theta_vcl}. These parameters allow us to estimate the molecular kinetic parameters for our system coated glass/PDMS/air as given by Eq.~(\ref{eq:v0_Gamma}). Thus, we find
\begin{equation}
 \lambda=5.274 \times 10^{-7}\text{cm},\qquad\kappa=2.171 \,\text{s}^{-1}.
\end{equation}
Interestingly, $\lambda$ is of the same order as those reported in~\cite{petrov_lang92}, but $\kappa$ is much smaller than for those systems with glycerol, likely due to the higher viscosity of PDMS. At the same time, $\ell$ is at least ten times larger here than in~\cite{petrov_lang92}.

\section{Dynamics of a liquid filament on the substrate: Experiments}
\label{sec:exp}

The main dynamical process that we analyze here is the axial dewetting motion of a liquid filament placed on a horizontal and previously coated glass substrate~\cite{gonzalez_04,gonzalez_07,rava_pof16}. We generate the filament from a vertical jet of PDMS flowing out from a small nozzle at the bottom of a vessel filled with PDMS. It is captured on the substrate by horizontally flipping the substrate in a rigid frame. Quickly afterwards, the frame is rotated $90$ degrees and a second horizontal flip is performed so that a new (auxiliary) filament, perpendicular to the first one, is also captured on the substrate. This other filament crosses the first one near one of its ends, so that the latter adopts a reproducible and characteristic rounded shape, whose axial dewetting will be observed and measured in detail. Finally, the substrate is rapidly placed on a horizontal position. All these movements take about $5$ seconds, which is a very short time interval compared to the time scale of the experiment itself due to the high viscosity of the PDMS. 

Thus, we obtain a fluid filament of uniform width, $w$, with parallel and straight contact lines, so that the initial configuration has a constant cross sectional area along its axis. We calibrate the system by relating the fluid height in the vessel and the jet diameter for a given nozzle. Both the jet diameter and the corresponding filament widths (for a contact angle equal to $\theta_a$) could be varied from $0.3$ to $1.3$~mm and from $0.1$ to $1.0$~mm, respectively.

\subsection{Axial dewetting between breakups}
\label{sec:fil_dewet}

A typical dewetting process of the filament end is shown in Fig.~\ref{fig:sequence}. Initially, it is rounded and has a contact angle of around $25^\circ$ so that the tip certainly recedes according to the wettability framework described above (see e.g. Fig.~\ref{fig:theta_vcl}). This motion induces the development of a head, whose width and thickness grow with time. The evolution proceeds until the filament tip stops and, after a while, a narrower region (neck) starts forming in the filament somewhere beyond the bulged region. Finally, the neck breaks up, so that the portion of fluid between the tip and the neck gives place to a static drop. The characteristic shape of this sessile drop with non-circular footprint was previously studied in~\cite{rava_pof16}, but here we will focus on the description of the dynamical process previous to the rupture. This dewetting and breakup mechanism repeats itself starting from a new end of the filament, and finishes with the formation of a series of similar drops. Here, we show results for the tracking of the first four heads at the left end of this filament.

\begin{figure}[thb]
\centering
\subfigure[ $w_a=0.082$~cm]
{\includegraphics[width=0.35\linewidth]{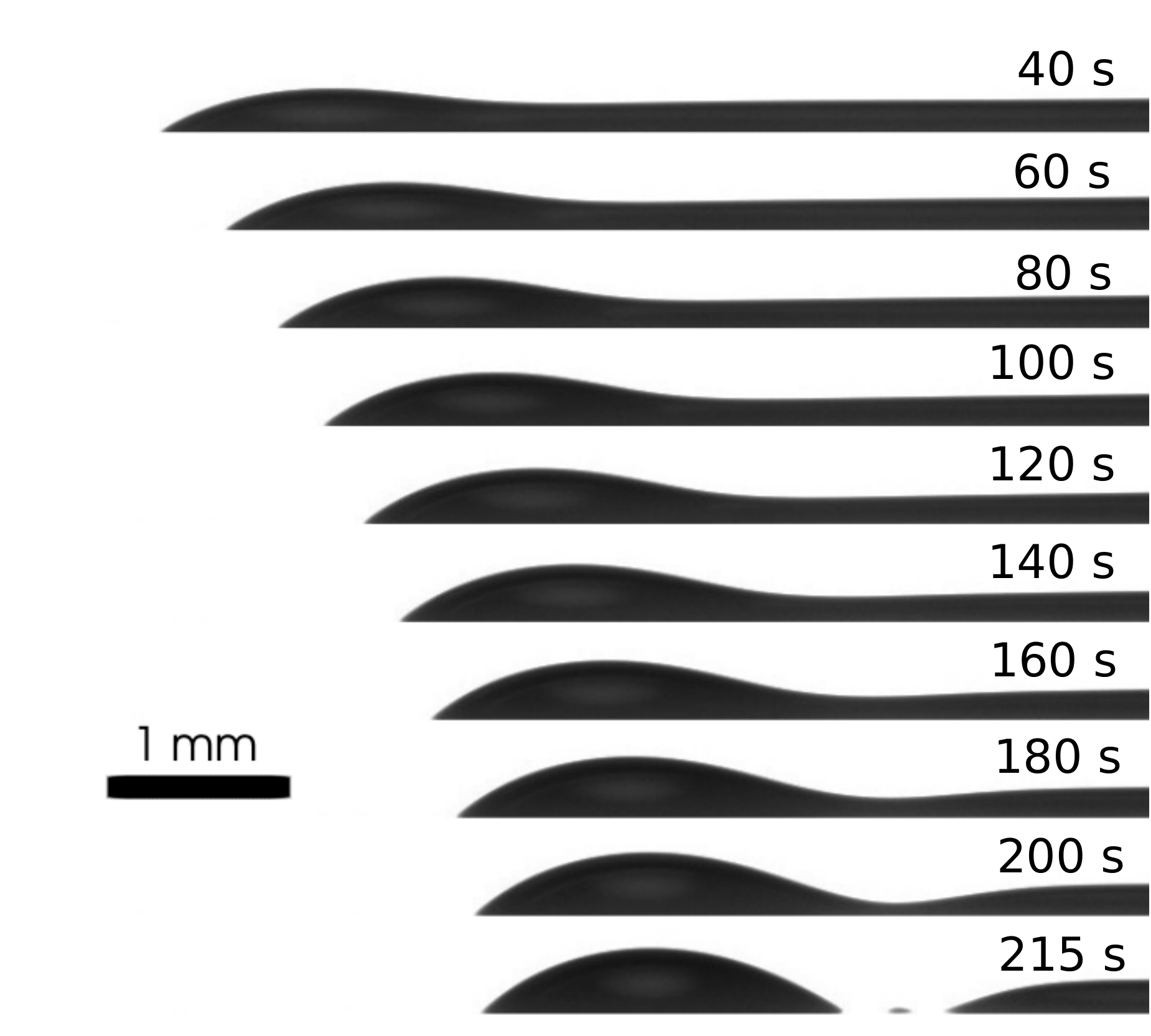}}
\hspace{0.2\linewidth}
\subfigure[ $w_b=0.0665$~cm]
{\includegraphics[width=0.16\linewidth]{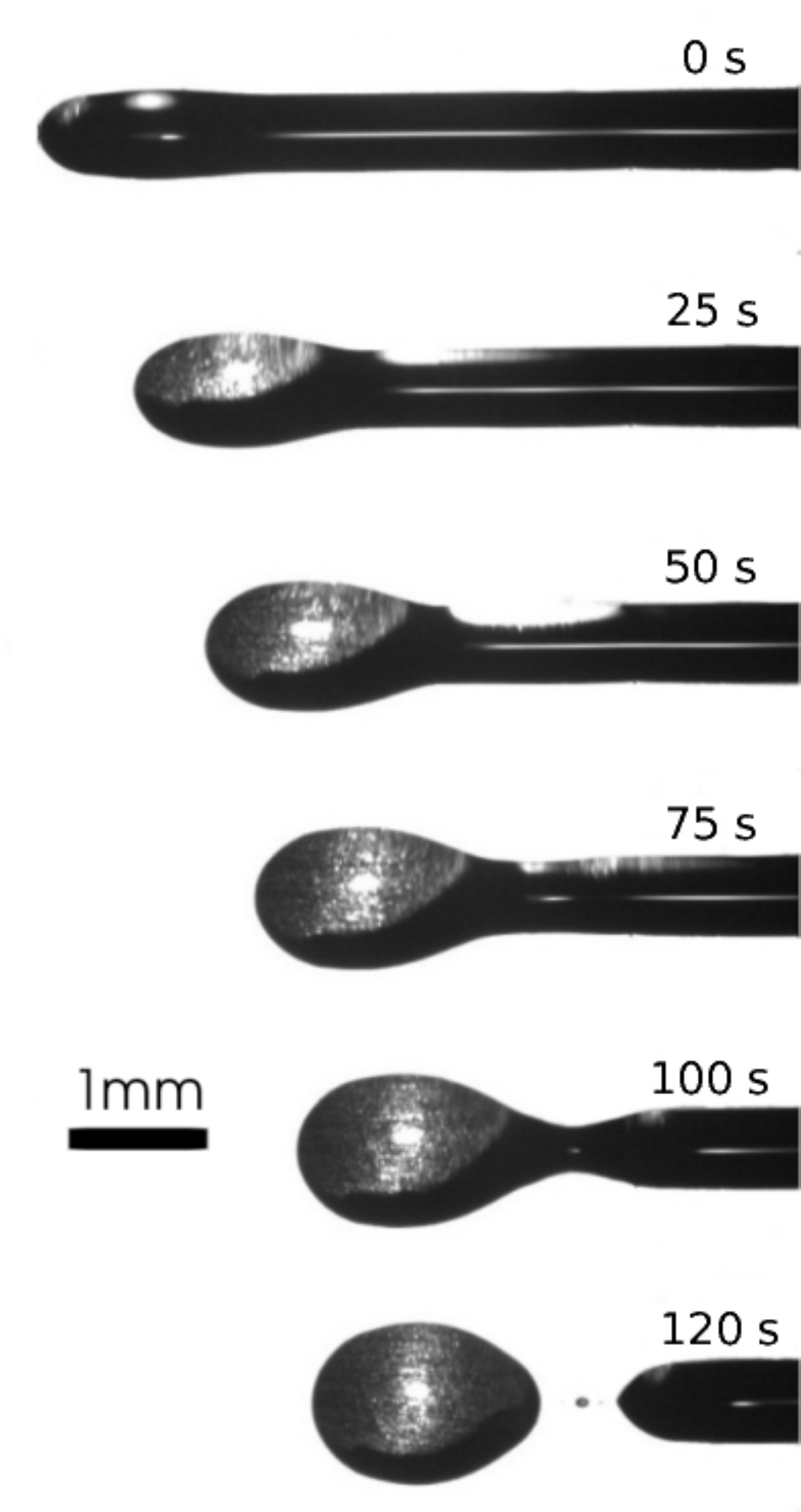}}
\caption{(a) Side and (b) top views of filaments with different width, $w$, at several times.}
\label{fig:side_top}
\end{figure}

The experimental setup allows to observe, though not simultaneously, both top and side views of the filament evolution. In Fig.~\ref{fig:side_top}, we show the filament profiles for two similar, albeit not strictly equal, widths. From this type of profiles, we are able to extract the position of the tip, the contact angle there, as well as the thickness and width of the head. These parameters are later compared with the results from an heuristic model and numerical simulations of the full hydrodynamic equations.

\subsection{Evolution of fronts, contact angles, and thicknesses}
\label{sec:fil_time}

It should be noted that the first drop formed from the end of the filament is a consequence of a different breakup process (numbered here as head $0$) than the following ones. This is due to the fact that its preceding breakup occurs at crossings with a transversal auxiliary filament, unlike the rest of the drops. Since the initial conditions of the formation process determines important features of the dynamics, such as the distance travelled by the retracing tip till the following breakup, we shall restrict ourselves to the most repetitive case. Therefore, we exclude it in our analysis.

\begin{figure}[hbt]
\centering
\subfigure[]
{\includegraphics[width=0.45\linewidth]{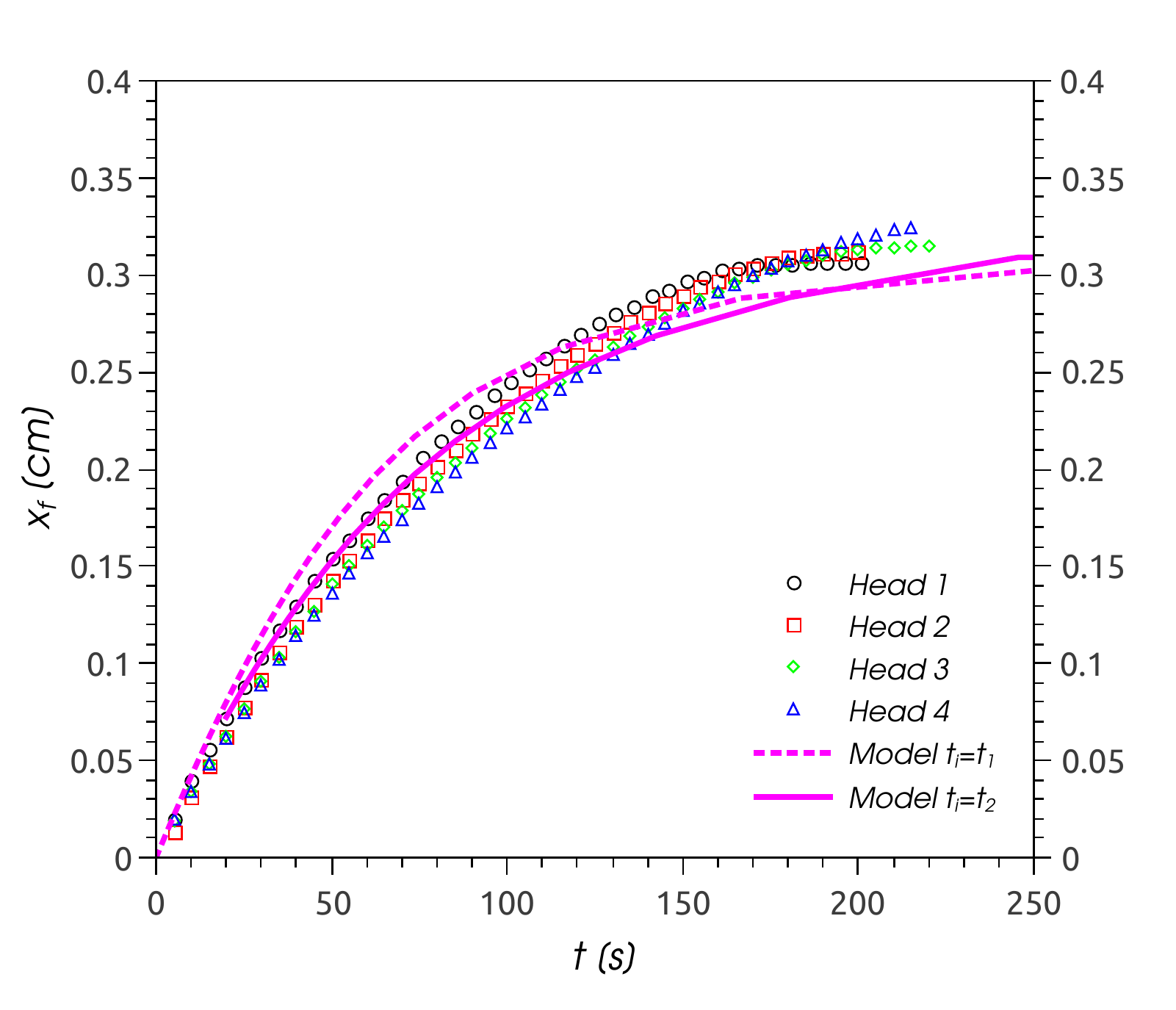}}
\subfigure[]
{\includegraphics[width=0.45\linewidth]{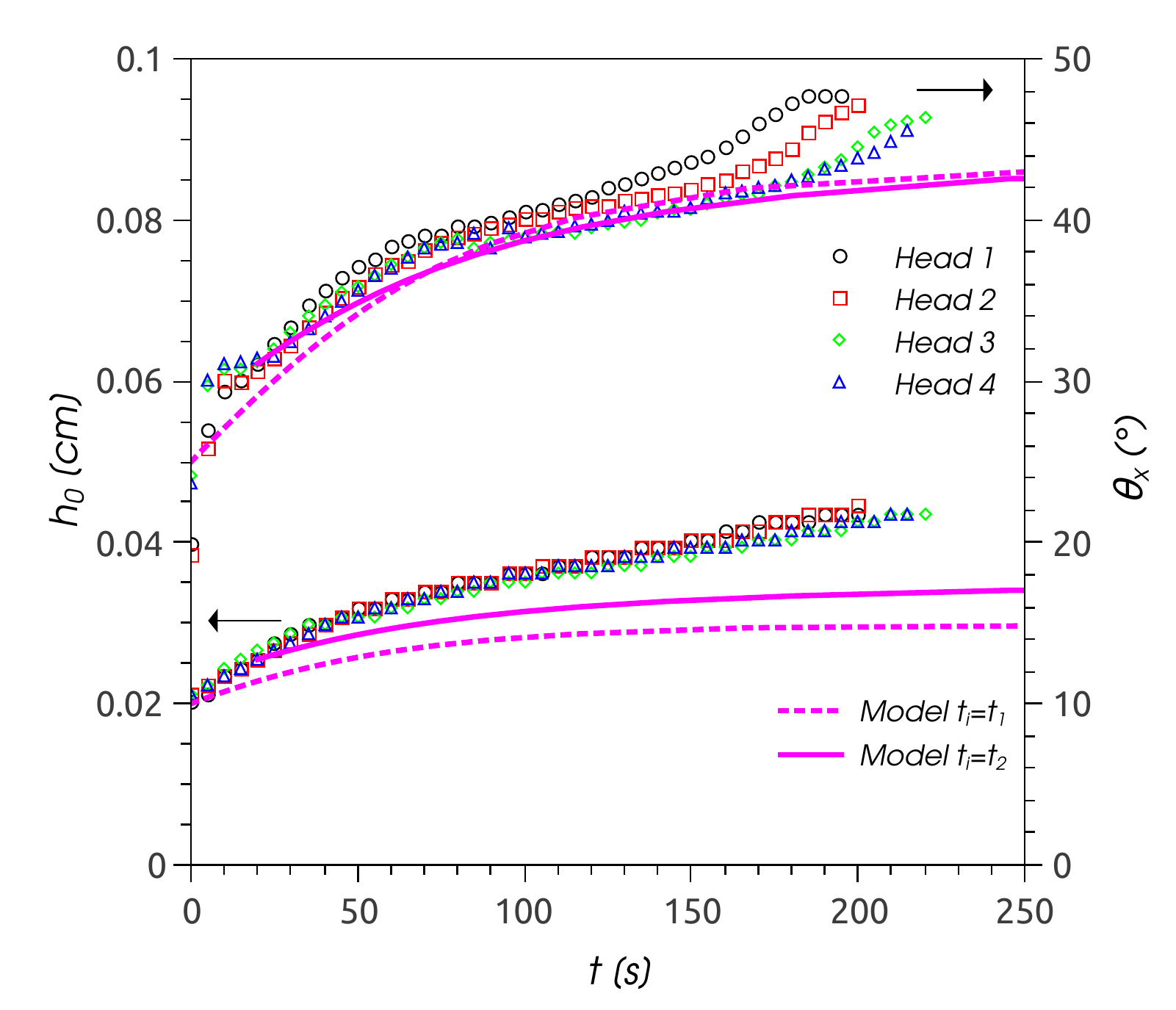}}
\caption{{\it Side view}. (a) Position, $x_f$, and (b) dynamic contact angle, $\theta_x$, and maximum thickness, $h_0$, versus time for the left end of a filament of width $w_a$. Every symbol corresponds to a different breakup, such that consecutive heads are formed. Thus, $t=0$ indicates the moment just after each breakup. The lines correspond to the results given by the model presented in Section~\ref{sec:model}.}
\label{fig:xf-thex-h0_17}
\end{figure}

In Fig.~\ref{fig:xf-thex-h0_17} we show with symbols the time evolution of the axial position of the tip (front), $x_f$, the dynamic contact angle there, $\theta_x$, and the maximum thickness of the head for the first four heads  ($1$ to $4$) formed at the left end of the filament of width $w_a$ (see Fig.~\ref{fig:side_top}(a)). The origin of $x_f$--coordinates for successive heads is the point where each breakup took place. The almost perfect superposition of the data for all three quantities ($x_f$, $\theta_x$, and $h_0$) indicates that the dewetting process is repeatable after each breakup. 

\begin{figure}[htb]
\centering
\includegraphics[width=0.45\linewidth]{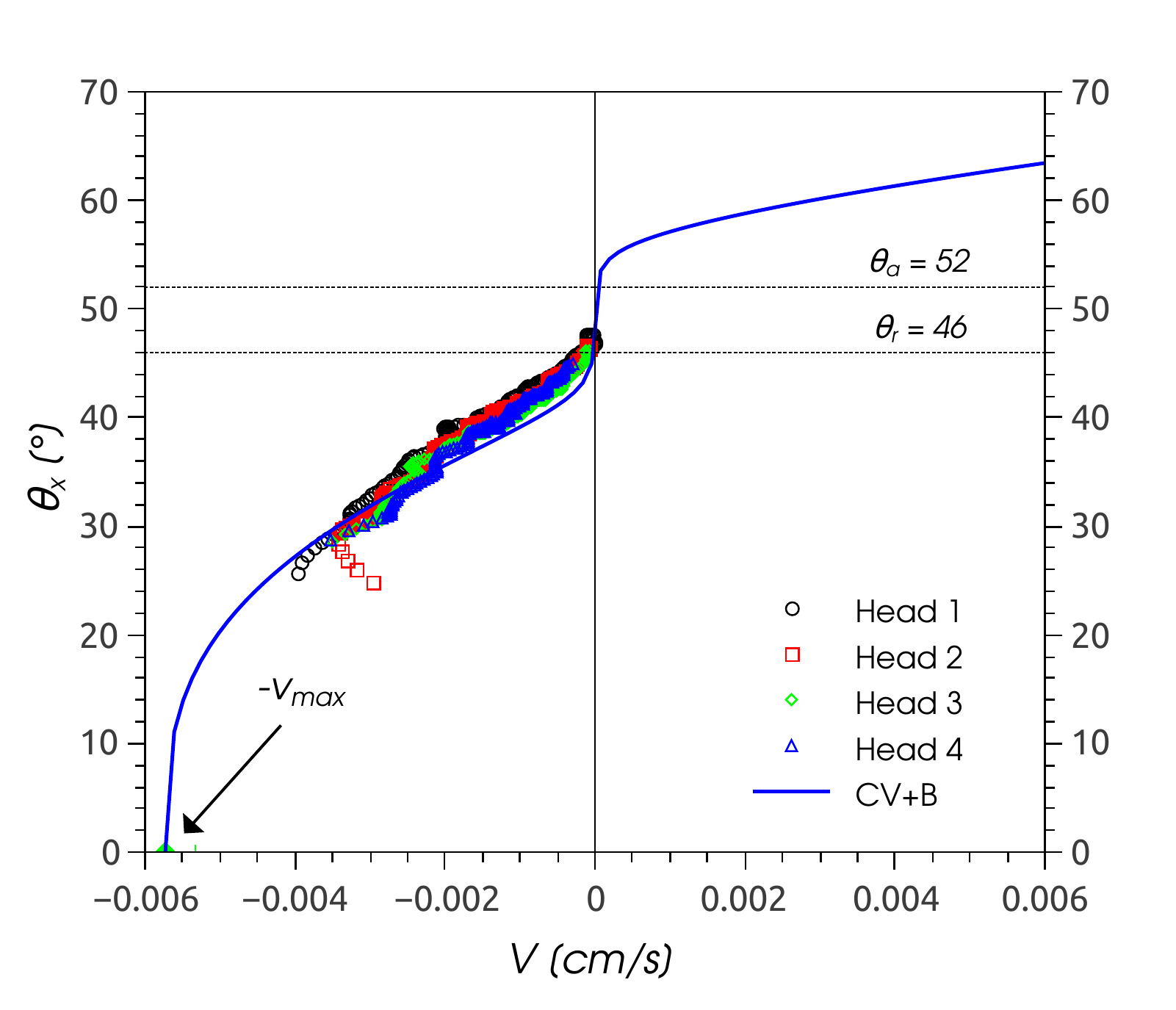}
\caption{Values of the dynamic contact angle at the tip of the receding filament, $\theta_x$, versus the contact line velocity, $v_{cl}$, as obtained from the data in Fig.~\ref{fig:xf-thex-h0_17}. The solid line corresponds to the {\it CV+B} relationship, Eq.~(\ref{eq:theta_fit}), with the parameters given in Eq.~(\ref{eq:param}). }
\label{fig:the_vcl-fil}
\end{figure}

From the data presented in Fig.~\ref{fig:xf-thex-h0_17}, we generate the corresponding $\theta_x$--$v_{cl}$ relation, as shown by symbols in Fig.~\ref{fig:the_vcl-fil}. Interestingly, these spontaneous filament dewetting data are in good agreement with the previously obtained relationship  $\theta(v_{cl})$ for the drop under the cyclic procedure of forced wetting and dewetting motions (solid line in Fig.~\ref{fig:the_vcl-fil}). This fact confirms that the previously obtained relationship certainly describes a local phenomenon, and is independent of the geometry of the flow. 

\begin{figure}[htb]
\centering
\subfigure[$w_b=0.0665$~cm]
{\includegraphics[width=0.45\linewidth]{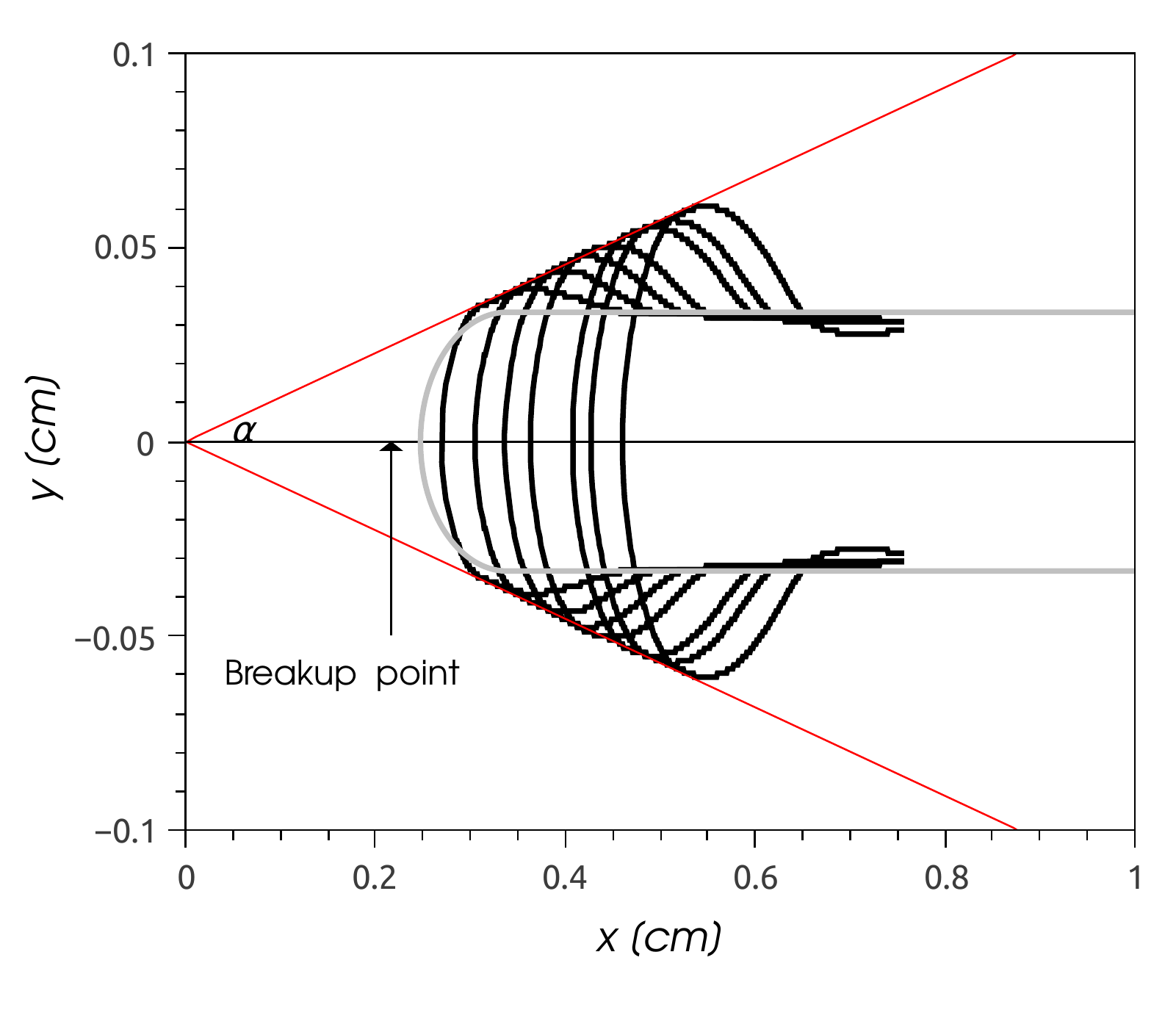}}
\subfigure[$w_c=0.107$~cm]
{\includegraphics[width=0.45\linewidth]{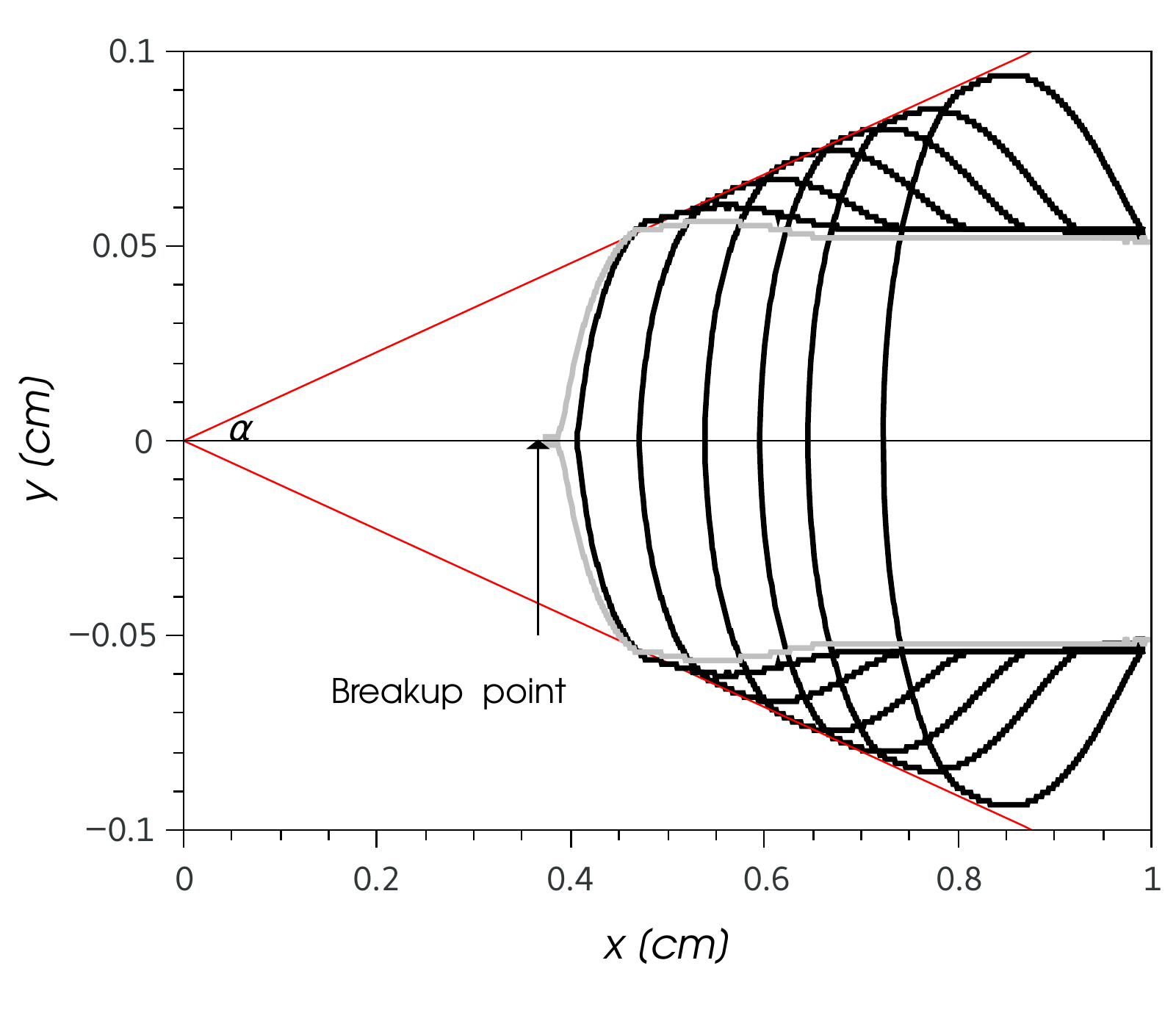}}
\caption{{\it Top view}. Contours (contact line shapes) for the filaments of width: (a) $w=w_a=0.0665$~cm (see Fig.~\ref{fig:side_top}(b)), and (b) $w=w_c=0.107$~cm. Note that the angle $\alpha=6.5^\circ$ is the same for both widths. The gray lines correspond to very early contours, just after the rupture and close to the breakup point. Note the spiked shape at the tip in case (b).}
\label{fig:conos}
\end{figure}

The measured evolution of the footprint of a filament can be seen in Fig.~\ref{fig:conos} for two different filament widths (see also  Fig.~\ref{fig:side_top}(b)). The results show that all footprints are inside an angular sector whose borders are always tangent at some point $(x_1,y_1)$ of the contact line. The aperture of this sector is fixed, so that the slopes of the straight lines are constant during the axial retraction process, except when the neck behind the head becomes very narrow. The straight lines satisfy the expression $y=\pm(0.114\,x +0.031)$~cm,  which means that they intersect at $(x_\ast,0) = (-0.2719,0)$~cm. We take this point as the origin of coordinates for the following discussions, so that this envelope can be described as straight lines with slopes $\alpha=6.5^\circ$. The physical meaning of this  result will become apparent from the analysis of the full simulations in Section~\ref{sec:num}. 

\begin{figure}
\centering
\subfigure[]
{\includegraphics[width=0.45\linewidth]{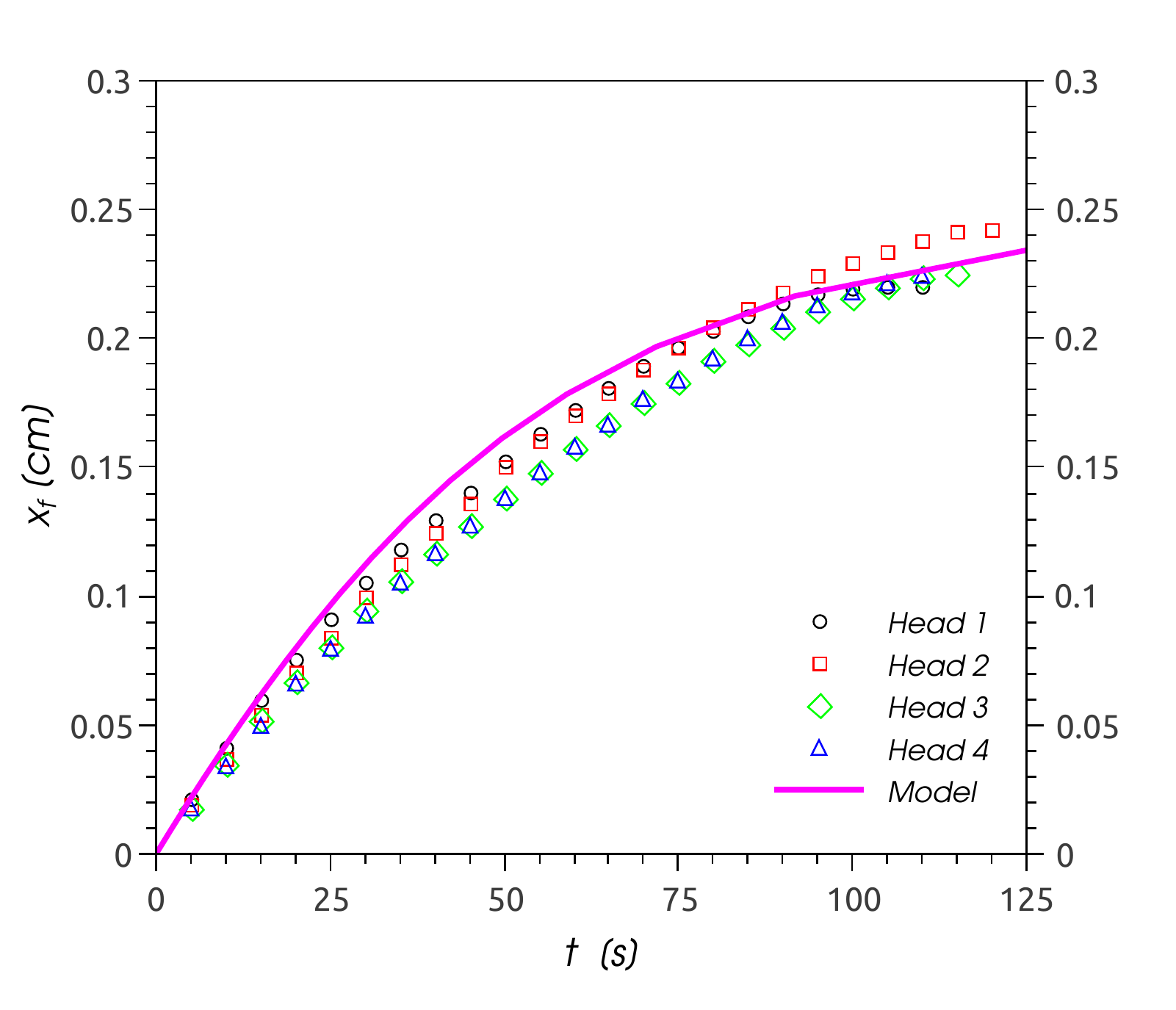}}
\subfigure[]
{\includegraphics[width=0.45\linewidth]{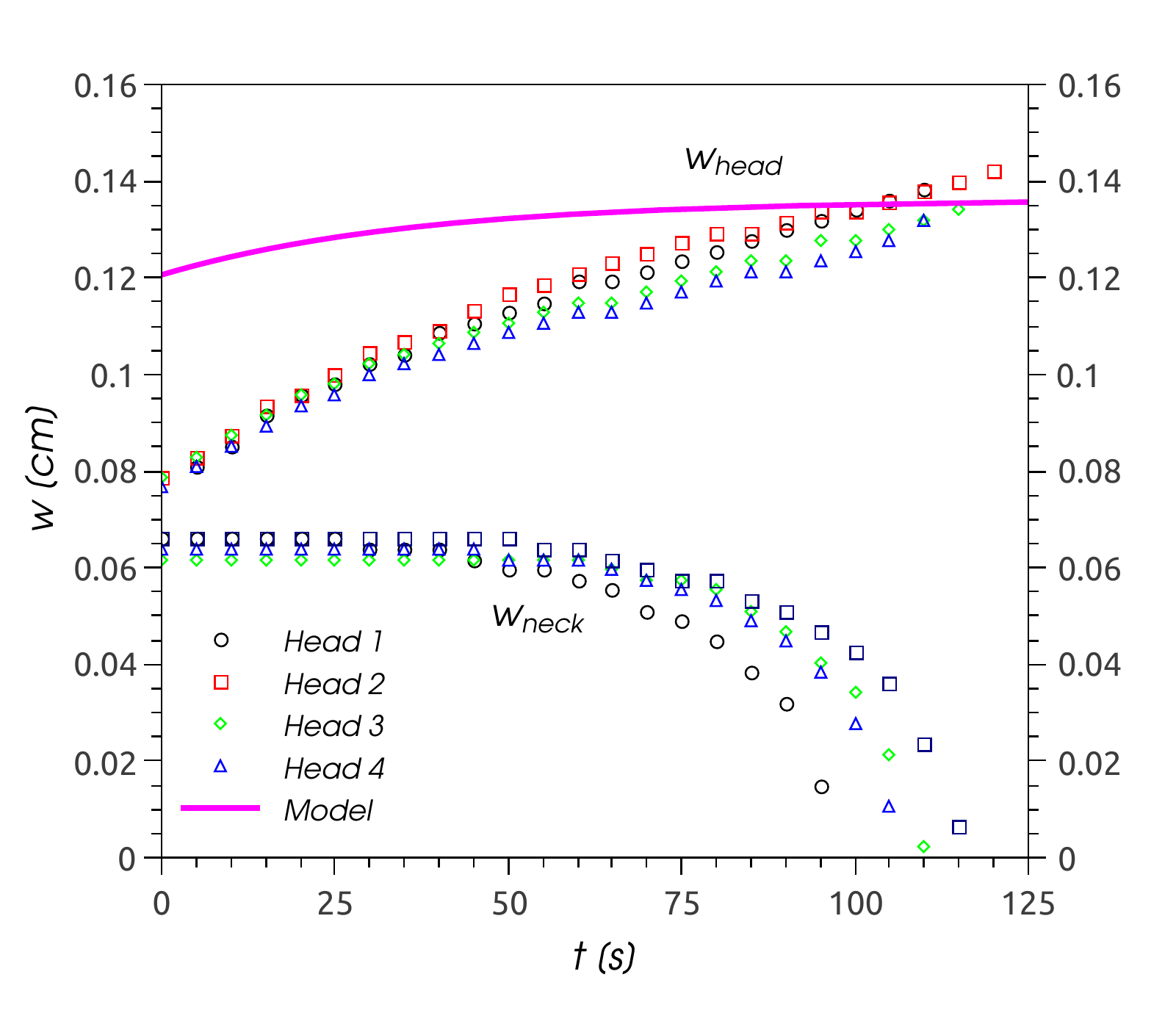}}
\caption{{\it Top view}. Experimental data (symbols) and numerical simulations (dashed lines) for a filament with $w_b$ observed from top (see Fig.~\ref{fig:side_top}(b)): (a) Position of the tip, $x_f$, and (b) maximum width at the head, $w_{head}$, and minimum width at the neck, $w_{neck}$.}
\label{fig:xf-w_20}
\end{figure}

In Fig.~\ref{fig:xf-w_20} we show the tip position, $x_f$, the maximum width at the head, $w_{head}$, and the minimum width at the neck, $w_{neck}$, as a function of time for the filament width $w_b=0.0665$~cm. As expected, the curve for $x_f$ is very similar to that in Fig.~\ref{fig:xf-thex-h0_17}(a). In particular, that for early times, say $t<100$~s, the data for both widths are practically coincident, the main difference being for later times when $x_f$ approaches the corresponding limiting value. On the other hand, Fig.~\ref{fig:xf-w_20}(b) shows $w_{head}$ and $w_{neck}$, which are obtained as the maximum and minimum values of the width along the filament. Note that the growth rate of the head slightly diminishes when the neck starts developing ($t \approx 40$~s).

\section{Description of the axial dewetting: Heuristic model}
\label{sec:model}

\begin{figure}[htb]
\centering
\includegraphics[width=0.65\linewidth]{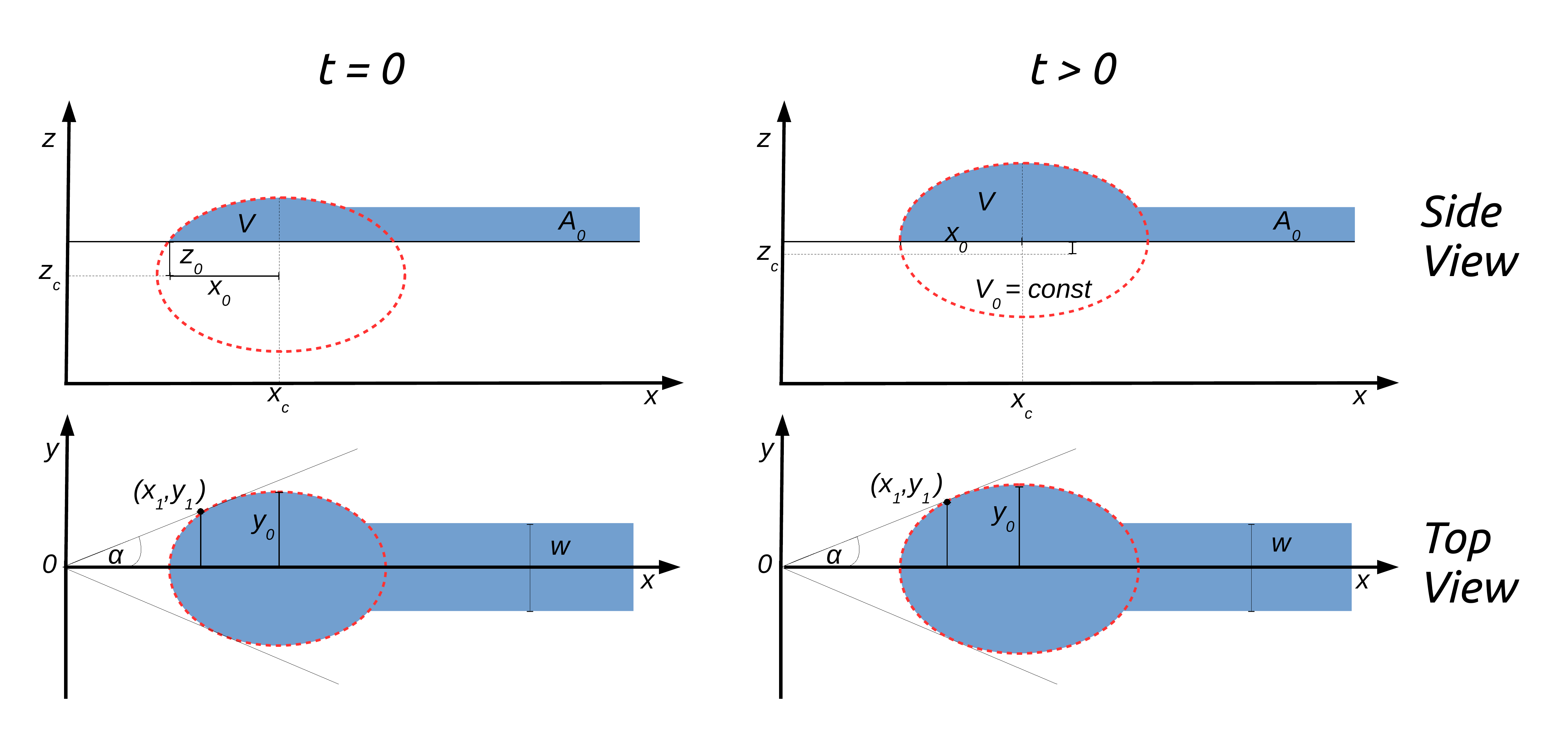}
\caption{Sketch showing how the model describes the shape and motion of the head at the extreme of a filament by means of a moving ellipsoidal cap. The upper line shows the side--view for $t=0$ and $t>0$, while the bottom line shows the top view for the same times.}
\label{fig:model}
\end{figure}

In order to develop a simple model that accounts for the main features of the dynamic process of dewetting, we consider that the shape of the head at the end of the filament can be approximated as a part of an ellipsoidal cap (see Fig.~\ref{fig:model}). This simple geometrical configuration continuously connects to the rest of the filament of circular cross section. The semi--diameters along $x$, $y$ and $z$ axes are $a$, $b$ and $c$, respectively, which we consider as constants during the whole evolution. The distance between the center of the ellipsoid and the substrate in the vertical direction is $z_0$, and the cap has length $2x_0$ along $x$-axis and width $2 y_0$ in the $y$ direction, as shown in Fig.~\ref{fig:model}(b).

The key idea is that the head can be emulated with the part of this ellipsoid that is above the substrate level. This cap volume, $V$, varies as the dewetting motion of the tip proceeds. The dynamics can be described as the superposition of both a vertical and a horizontal motion of the ellipsoid, with velocities $dz_0/dt$ and $U$, respectively. Since the ellipsoid is not only emerging but also moving horizontally, we describe it in terms of the coordinates of its center,  $(x_c,0,z_c)$, and thus, it is described by the equation 
\begin{equation}
\frac{(x-x_c)^2}{a^2}+\frac{y^2}{b^2}+\frac{(z-z_c)^2}{c^2}=1.
\label{eq:elipsoid}
\end{equation}
Then, the volume of the ellipsoidal cap is given by
\begin{equation}
	V(z_0)= \pi a b \left( \frac{2c}{3} - z_0 + \frac{z_0^3}{3 c^2} \right).
\end{equation}
Due to the horizontal retraction of the tip, there is a mass accretion in the cap from the filament that is being swept by the receding head. Therefore, the cap volume $V(z_0)$ increase can be calculated in first approximation by
\begin{equation}
	\frac{dV}{dt}= A_0 U,
	\label{eq:dVdt}
\end{equation}
where $A_0$, the area of the cross section of the filament of width $w$ and transverse contact angle $\theta_a$, is given by
\begin{equation}
A_0= \left(\frac{w} {2 \sin \theta_a}\right)^2 \left( \theta_a - \frac{\sin 2\theta_a}{2} \right) .
\end{equation}

Here, we define $x_0$ as the distance between the tip and the center position, $x_c$, i.e. $x_0=x_f-x_c$. Note that the vertical motion of the ellipsoid yields a variation of $x_0$. Therefore, the contact line velocity along $x$--axis is the result of the superposition of the bulk horizontal displacement and the variation of $x_0$,
\begin{equation}
 v_f= \frac{dx_0}{dt} + U,
 \label{eq:vel}
\end{equation}
where $U=dx_c/dt$ and $v_f=v_{cl}(\theta_x)=dx_f/dt$ is the front (tip) velocity as a function of the dynamic contact angle at the tip, $\theta_x$,  whose functional form is given implicitly by Eq.~(\ref{eq:theta_fit}). The vertical motion of the ellipsoid and the horizontal displacement are not independent since
\begin{equation}
 \frac{dz_0}{dx_0}=-\tan \theta_x,
\end{equation}
so that there is an increase of the volume in the head as well as a variation of the contact angle, $\theta_x$. Upon writing $dV/dt=V'(z_0) dz_0/dt$, Eqs.~(\ref{eq:dVdt}) and (\ref{eq:vel}) lead to
\begin{equation}
 \frac{dx_0}{dt} = \frac{A_0 v_f(\theta_x)}{A_0 - V'(z_0) \tan \theta_x },
 \label{eq:dx0dt}
\end{equation}
whose integration allows a comparison with the time evolution of most of the parameters measured in the experiments.

In order to find the values of the constants $a$ and $c$ of the ellipsoid, we consider Eq.~(\ref{eq:elipsoid}) at the tip position ($y=0$),
\begin{equation}
 \frac{x_0^2}{a^2}+\frac{z_0^2}{c^2}=1,
 \label{eq:eqtip}
\end{equation}
and its slope (contact angle) 
\begin{equation}
 \tan \theta_x = - \frac{dz_0}{dx_0} = \frac{x_0 c^2}{z_0 a^2},
 \label{eq:thex}
\end{equation}
where 
\begin{equation}
 x_0=a \sqrt{1-(z_0/c)^2}.
 \label{eq:x0}
\end{equation}
For given values of $x_0$, $\theta_x$, and $h_0$ ($=c-z_0$), Eqs.~(\ref{eq:eqtip})--(\ref{eq:x0}) yield, 
\begin{equation}
 a= \frac{x_0}{2} \sqrt{\frac{4 \xi - \xi^2 \cot\theta_x - 4\tan\theta_x}{\xi-\tan\theta_x}}, \qquad 
 c=\frac{h_0}{2}\frac{\xi-2 \tan\theta_x}{\xi-\tan\theta_x},
 \label{eq:ac_side}
\end{equation}
where $\xi=2 h_0/x_0$ must satisfy the condition $\xi<\tan\theta_x$. Here, we consider $x_0$, $\theta_x$, and $h_0$ at $t=0$.

On the other hand, to calculate $b$ we consider Eq.~(\ref{eq:elipsoid}) at the point of the contact line where the width of the head is maximum ($x=x_c$),
\begin{equation}
 \frac{y_0^2}{b^2}+\frac{z_0^2}{c^2}=1,
 \label{eq:eqwmax}
\end{equation}
and its slope (contact angle) 
\begin{equation}
 \tan \theta_y = - \frac{dz_0}{dy_0} = \frac{y_0 c^2}{z_0 b^2},
 \label{eq:they}
\end{equation}
where
\begin{equation}
 y_0=b\sqrt{1-(z_0/c)^2}.
 \label{eq:y0}
\end{equation}

By using Eqs.~(\ref{eq:eqtip})--(\ref{eq:x0}) and (\ref{eq:eqwmax})--(\ref{eq:y0}), we define the following ratios:
\begin{equation}
\phi  \equiv \frac{b}{a} = \frac{\tan \theta_x}{\tan \theta_y}=\frac{y_0}{x_0}.
\label{eq:y0x0}
\end{equation}
Therefore, in order to determine $b$ in terms of $a$, any of these equalities could be used. Since $\theta_x$ and $\theta_y$ cannot be measured simultaneously, we discard the ratio of their tangents. Moreover, even if one could attempt to produce an $yz$--plane view, any measurement of $\theta_y$ would be blocked by the drops of the previous ruptures. Instead, we consider the ratio $y_0/x_0$, which can be measured from top view observations.  From experiments with three different widths ($w_b$, $w_c$ and $w_d=0.128$~cm) and four ruptures for each width, we obtain an average value of the ratio $x_0/y_0$, as $\phi=0.599 \pm 0.08$.

Finally, we obtain the function $v_f(\theta_x)$ inverting $\theta(v_{cl})$ evaluated at the tip, see Eq.~(\ref{eq:theta_fit}), and  solve Eq.~(\ref{eq:dx0dt}) numerically in time. Thus, we have 
\begin{equation}
 t=\int_{x_{0,i}}^{x_0(t)} \frac{A_0-V'(z_0) \tan\theta_x(x_0)}{A_0 v_f(\theta_x(x_0))} \,dx_0,
\end{equation}
where $\theta_x(x_0)$ is given by Eq.~(\ref{eq:thex}), and $x_{0,i}=x_0(t=0)$. Once $x_0$ versus $t$ is found, we can calculate the contact angle $\theta_x(x_0(t))$, the maximum thickness $h_0(t)$, and the tip position
\begin{equation}
 x_f(t)=x_0(t) - x_{0,i} + \int_{x_{0,i}}^{x_0(t)} \frac{V'(z_0)}{A_0}  \tan \theta_x \, dx_0= x_0(t) - x_{0,i} + I(x_0(t)),
 \label{eq:xft}
\end{equation}
where the integral $I(x_0(t))$ is
\begin{equation}
 I(x_0(t))=   \frac{\pi b c}{A_0 a^3} \int_{x_{0,i}}^{x_0(t)} \frac{x_0^3}{\sqrt{1-(x_0/a)^2}} \, dx_0
  = \frac{\pi b c}{3 a A_0} \left[ \left( 2 a^2 + x_0^2 \right)\sqrt{1-(x_0/a)^2}\right]_{x_{0,i}}^{x_0(t)}.
  \label{eq:It}
\end{equation}

We show in Fig.~\ref{fig:xf-thex-h0_17} a comparison of the time evolution as predicted by the model with the experimental data. As a first attempt, we determine the constants $a$, $b$, and $c$ by taking the values of $x_0$, $h_0$, and $\theta_x$ corresponding to the first frame capture after the breakup ($t=0$). Thus, we obtain the curves shown as dashed magenta lines in Fig.~\ref{fig:xf-thex-h0_17} ($t_i=t_1$). In spite of the simplicity of the model, based on a rough approximation of the free surface shape (assumed as an ellipsoidal cap), the agreement with experiments is remarkably good, except for the maximum thickness of the head, $h_0(t)$. Note, however, that immediately after the breakup the shape of the head is strongly changing from a spike to a rounded contour (see Fig.~\ref{fig:conos}). Thus, during a short time of the order of a few seconds, the head does not achieve an ellipsoidal shape, and then, a too early selection of the initial values of $x_0$, $h_0$, and $\theta_x$ can lead to an inaccurate determination of $a$, $b$, and $c$. Moreover, since these initial profiles are quite flat at the top, with $h_0$ being close to the filament thickness, it is difficult to determine $x_0$ with sufficient accuracy.

Better agreement with experiments could be expected if the  initial values are taken from a later head whose shape has settled closer to that of an ellipsoidal cap. Therefore, we use instead the second frame ($\approx 20$~s later) as starting time, and obtain new results from the model (see solid magenta lines in Fig.~\ref{fig:xf-thex-h0_17}, $t_i=t_2$). We observe that the agreement with experiments is greatly improved for $x_f(t)$ and $\theta_x(t)$, and even the $h_0(t)$ curve better approximates the experimental data.

Another interesting feature of the flow that can be extracted from the model is the maximum displacement of the receding tip, $x_{f,max}$. In fact, according Eq.~(\ref{eq:theta_fit}), the tip stops when $\theta_x$ reaches the value $\theta_0$. Thus, the maximum value of $x_0$ is given by  Eq.~(\ref{eq:thex}) in the form
\begin{equation}
 x_{0,max}=\frac{a^2 \tan \theta_0}{\sqrt{c^2+a^2 \tan^2 \theta_0}},
\end{equation}
which yields a maximum value for the tip position, $x_{f,max}$, as given by Eqs.~(\ref{eq:xft}) and (\ref{eq:It}) for $x_0=x_{0,max}$. The values of $x_{f,max}$ for the first four heads in Fig.~\ref{fig:xf-thex-h0_17} are shown in Table~\ref{tab:xfmax}. We note that the ratio $F=x_{f,max}/w$ is practically constant in experiments. The corresponding theoretical $F$ as obtained from our simple model is $F_{theo}= 3.82$, which is in good agreement with the average experimental value. This result is in accordance with one of the features of the model, namely that all the distance parameters in Eq.~(\ref{eq:xft}), such as $x_0$, $h_0$, $a$, $b$ and $c$, are proportional to $w$.

\begin{table}[htb]
	\begin{tabular}{ccr}\hline 
	Head & $x_{f,max}$ (cm) &$F_{exp}$   \\ \hline 
	$1$  & $0.306$ &$3.73$          \\ 
        $2$  & $0.312$ &$3.80$          \\ 
        $3$  & $0.315$ &$3.84$          \\ 
        $4$  & $0.324$ &$3.95$         \\ \hline 
	\end{tabular} 
	\caption{Experimental values of the ratio $F=x_{f,max}/w$ for the four heads in Fig.~\ref{fig:xf-thex-h0_17} with $w=w_a=0.082$~cm. The experimental average value  $F_{exp}= 3.83$, while the model yields $F_{theo}=3.82$}
	\label{tab:xfmax}
\end{table}

The use of the model for top view experiments is a bit more involved due to the lack of information about the contact angle at the tip, which is essential to the dynamics of the system. However, taking into account the proportionality of the coefficients with $w$ (see Table~\ref{tab:xfmax}), it is possible to infer the values of the semi--axes of the ellipsoid by using the information from a single side view experiment, such as the case above. Thus, by multiplying the previous values of $a$, $b$, and $c$ by the ratio $w_a/w_b$, we find the model results shown in Fig.~\ref{fig:xf-w_20}. Even with this rather indirect method to determine the parameters, the model is able to predict reasonably well the evolution of the tip position, $x_f(t)$. Instead, it fails to properly describe $w_{max}(t)$, and yields higher values than expected. This result is consistent with the fact that $h_0(t)$ is underestimated in the side view case (see Fig~\ref{fig:xf-thex-h0_17}(b)), since it implies an increase of $w_{max}$ to balance the mass flow swept by the receding tip. Note that the top view experiments do not provide any out of plane information, and therefore the input to the model dynamics is really scarce, so that the fact that it can pretty well describe at least $x_f(t)$ is remarkable.

These results suggest that the shape of the head can be roughly considered an ellipsoidal cap during most the receding stage. However, our model does not include any information about the transversal wetting, so it does not have the necessary ingredients to accurately predict the evolution of its width, $w_{max}(t)$, and maximum thickness, $h_0(t)$. On the other hand, the experimental evidence shows that by considering $x_0$, $h_0$, and $\theta_x$ at successive times, the corresponding values of $a$, $b$, and $c$ vary with time. In fact, while $a$ slightly oscillates around a typical value (so that one could assume an average constant), $b$ and $c$ steadily increase with time. Thus, the differences in $h_0(t)$ observed between the model of constant semi-axes and the experiments can be explained in terms of the increasing length of transverse semi-axes of the actual approximating ellipsoid.

\section{Numerical simulations}
\label{sec:num}

\subsection{Basic equations and method}
We numerically simulate the evolution of one end of the filament, and assume that the other one is so far away that it does not affect the other. Thus, we consider that at $t=0$ the filament starts at $x=0$ with the shape of a cylindrical cap of length $L_0$, width $w$ ($\ll L_0$), and with transverse equilibrium contact angle, $\theta_a$, along both parallel contact lines. The fluid region around the origin ($x \gtrsim 0$, $|y|\leqslant w/2$) remains practically at rest, while the main flow develops far away from there, close to the end region ($x \lesssim L_0$). Therefore, the boundary condition at $x=0$ is that for a no-slip wall. 

\begin{figure}
\centering
\subfigure[$t=0$]
{\includegraphics[width=0.245\linewidth, angle=180]{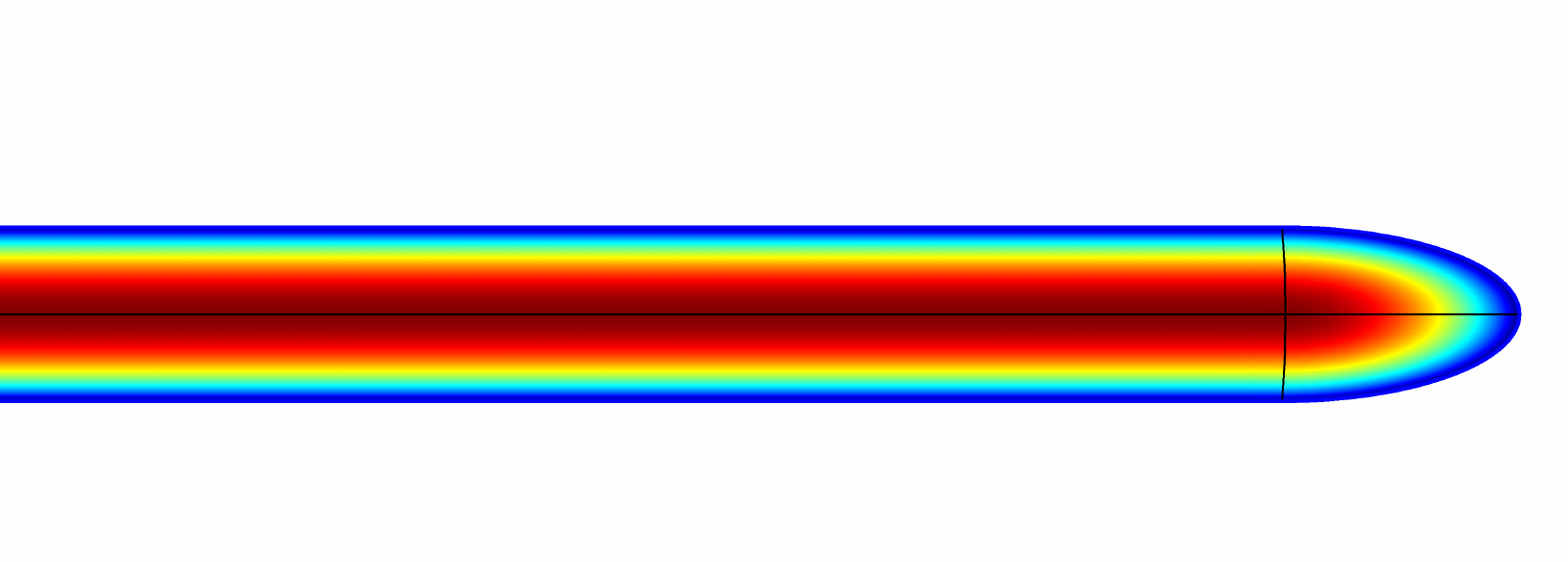}}
\subfigure[$t=500 t_c$]
{\includegraphics[width=0.245\linewidth, angle=180]{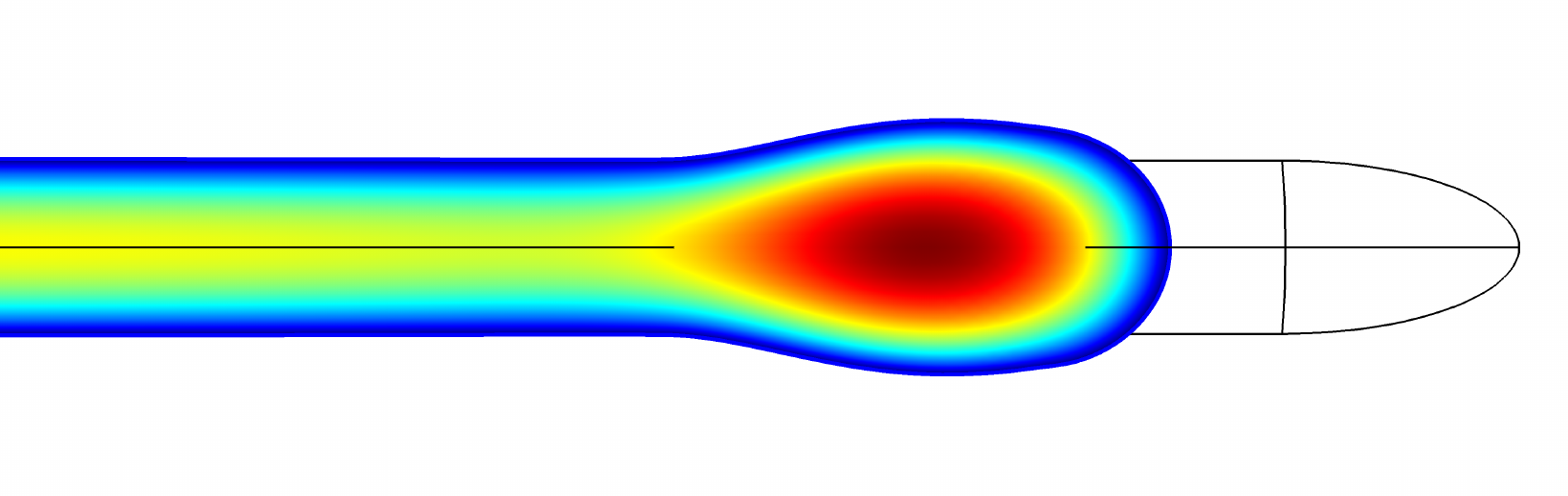}}
\subfigure[$t=1000 t_c$]
{\includegraphics[width=0.245\linewidth, angle=180]{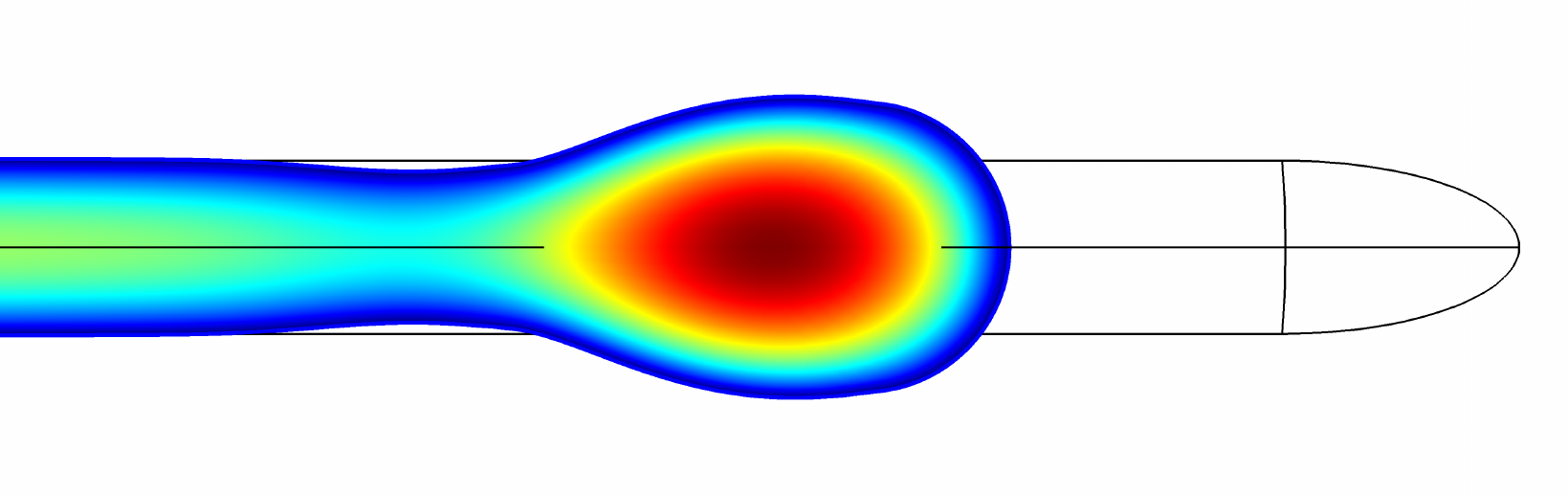}}
\subfigure[$t=1350 t_c$]
{\includegraphics[width=0.245\linewidth, angle=180]{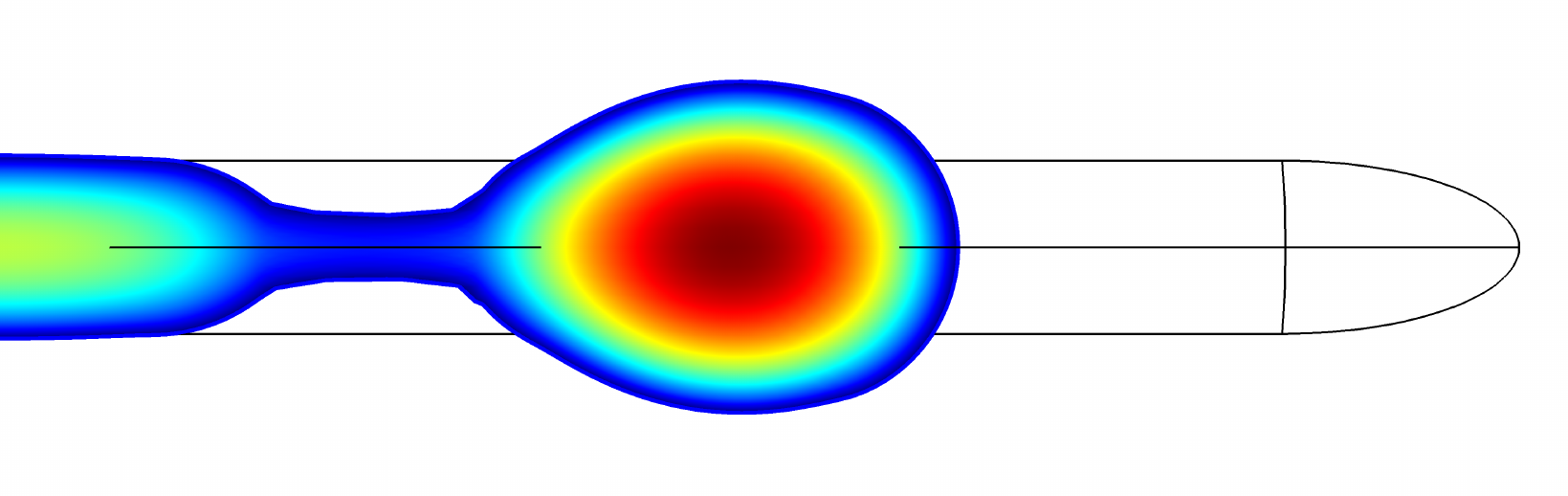}}
\caption{Time evolution of the filament of width $w_a=0.552 a_c$. The initially cylindrical part of the filament has a length $L= 8 a_c$, and lateral contact angle $\theta_a=52^\circ$. The initially ellipsoidal nose has contact angle $\theta_{x,i}=25^\circ$ at the tip. The time scale is  $t_c=\mu a_c/\gamma=0.1536$~s.}
\label{fig:3d}
\end{figure}

In order to emulate the actual filament, which ends at a rounded shape due to the breakup process that took place before ($t<0$), we here assume that this shaped end can be approximated by an additional cap of length $x_0$ at the end of the cylinder ($x=L$). Thus, the fluid domain is composed by a cylindrical cap of length $L_0$ plus an ellipsoidal cap of length $x_0$. The length $x_0$ is chosen in such a way that the filament tip has a given contact angle $\theta_{x,i}$, resulting from the breakup stage (see Fig.~\ref{fig:3d}(a)). 

The time evolution of this liquid filament of total length $L=L_0+x_0$ is obtained by numerically solving the dimensionless Navier-Stokes equation
\begin{equation}
La \left[ \frac{\partial \vec {v}}{\partial t}+(\vec{v} \cdot \vec {\nabla} ) \cdot \vec{v} \right] 
= - \vec {\nabla} p + \nabla ^{2} \vec {v} -\vec{z},
\label{eq:NS}
\end{equation}
where the last term stands for the gravity force. Here, the scales for the position $\vec x=(x,y,z)$, time $t$, velocity $\vec {v}=(u,v,w)$, and pressure $p$ are the capillary length $a_c$, $t_c=\mu a_c/\gamma$,$\gamma/\mu$, and $\gamma/a_c$, respectively. Therefore, the Laplace number is $La=\rho \gamma a_c/\mu^2$. In our experiments we have $a_c=1.49$ mm and $La=0.006$, so that inertial effects are practically irrelevant.  The $x$ and $y$--axes are assigned along and across the original filament, respectively.  Besides, the normal stress at the free surface accounts for the Laplace pressure in the form
\begin{equation}
 \Sigma_n = -\left( \vec \nabla_{\tau} \cdot \hat n \right) \hat n,
\end{equation}
where $\hat n=(n_x,n_y,n_z)$ and $\hat \tau$ are the versors standing for the normal and tangential directions to the free surface. Since, the surrounding fluid (e.g., air) is passive, we assume that the tangential stress is zero, i.e. $\Sigma_{\tau}=0$.

As regards to the boundary condition at the contact line, the dynamic contact angle, $\theta$, is given by the dimensionless contact line velocity, $Ca=\mu v_{cl}/\gamma$, according to the CV+B model in the form (see Eq.~(\ref{eq:theta_hyd}))
\begin{equation}
 \theta^3 = \arccos^3 \left[ \cos \theta_0 - \frac{1}{\Gamma} \sinh^{-1} \left( \frac{Ca}{Ca_0} \right) \right] + 
 9 Ca \ln \left( \frac{1}{\hat \ell} \right),
 \label{eq:theta_fit_num}
\end{equation}
where $Ca_0=\mu v_0/\gamma$, and $\hat \ell = \ell/a_c$. The contact line velocity is calculated from the velocity field as $Ca= N_x u + N_y v$, where $(N_x,N_y)=(n_x,n_y)/\sqrt{n_x^2+n_y^2}$ is the versor normal to the contact line. Note that this condition introduces a high nonlinearity to the problem, since the solution itself, namely the velocity field at $z=0$, yields the corresponding contact angle. 

We use a Finite Element technique in a domain which deforms with the moving fluid interface by using the Arbitrary Lagrangian-Eulerian (ALE) formulation~\cite{hughes_cmame81,donea_cmame82,christodoulou_cmame92,hirt_cmame97}. The interface displacement is smoothly propagated throughout the domain mesh using the Winslow smoothing algorithm~\cite{winslow_jcp66,knupp_ec99}. The main advantage of this technique compared to others such as the Level Set or Phase Field techniques is that the fluid interface is and remains sharp~\cite{tezduyar_cmame06}. The main drawback, on the other hand, is that the mesh connectivity must remain the same, which precludes the modeling of situations for which the topology might change (rupture of the filament). The default mesh used throughout is unstructured, and typically has $3\times 10^4$ triangular elements (linear elements for both velocity and pressure).  The mesh nodes are constrained to the plane of the boundary they belong to for all but the free surface.

\subsection{Comparison with experiments}

In order to compare the numerical results with the experiments, we perform simulations with the wettability parameters as given by Eq.~(\ref{eq:param}). Although the calculation cannot be continued beyond the breakup moment, the simulation is useful enough to account for the dewetting and wetting processes between breakups. Thus, the initial condition used here tries to emulate the scenario just after an actual breakup, and the resulting fluid motion is simulated till just before the next breakup. Fig.~\ref{fig:3d} shows the time evolution of the filament of width $w_a$. Here, we use $L_0= 8$ to define the cylinder length (we have checked that for $L_0> 6$, the results do not depend on $L$ at all), and $\theta_{x,i}=25^\circ$ at the tip, which determines the ellipsoidal shape that emulates the filament end (or nose) just after the rupture.

\begin{figure}[hbt]
\centering
\subfigure[]
{\includegraphics[width=0.45\linewidth]{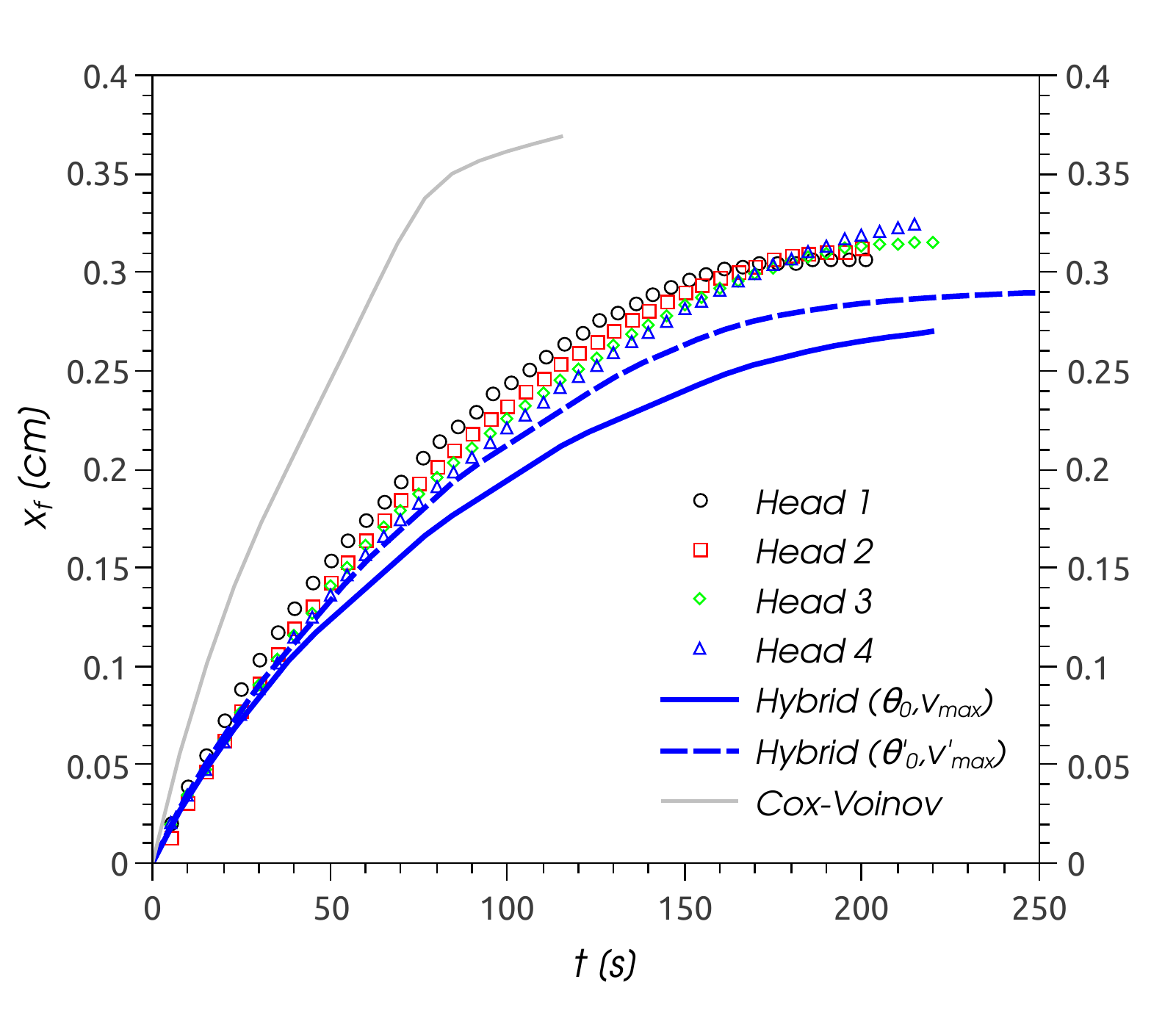}}
\subfigure[]
{\includegraphics[width=0.45\linewidth]{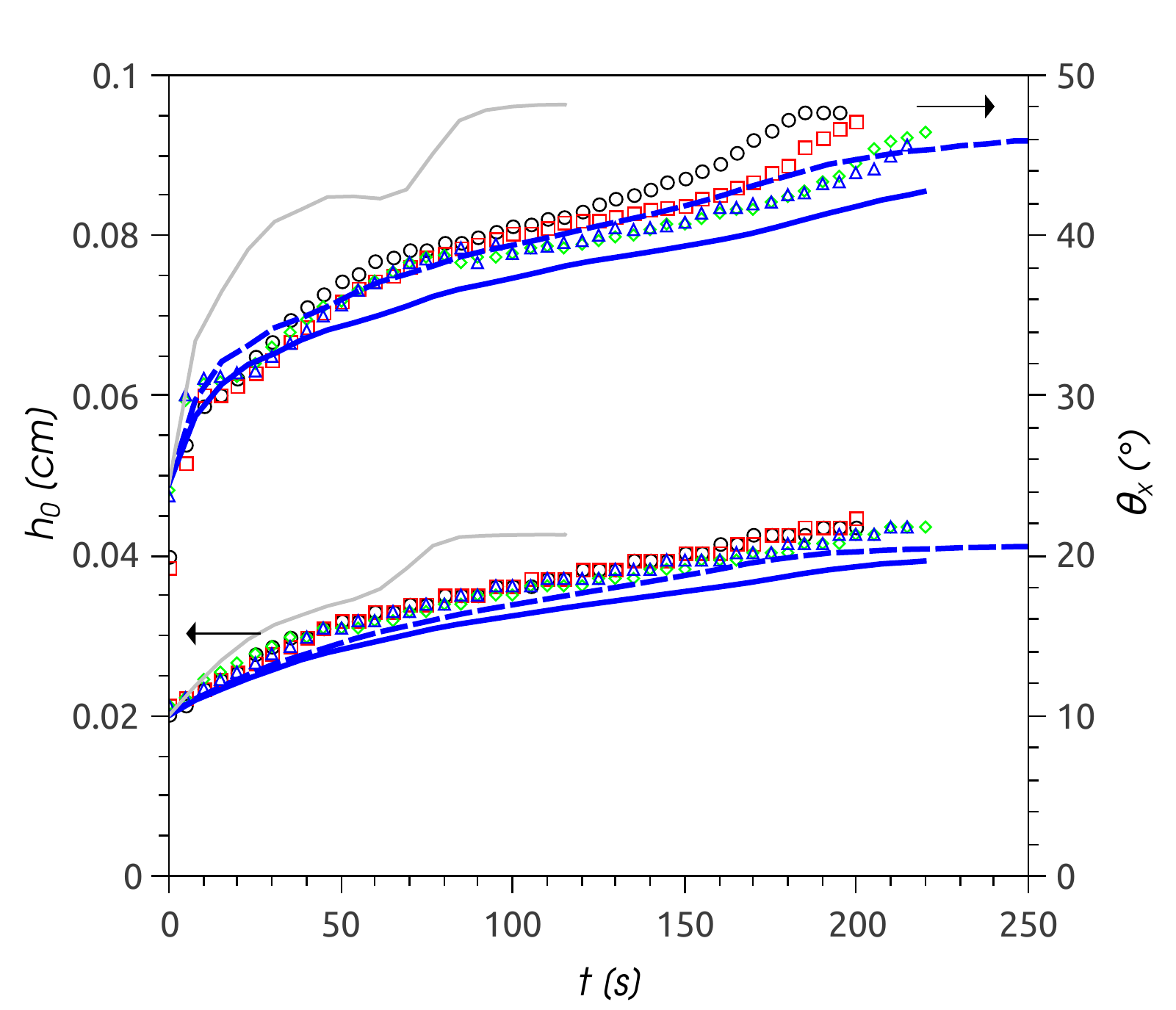}}
\caption{{\it Side view}. Comparison of the numerical solution with the experimental data in Fig.~\ref{fig:xf-thex-h0_17} (symbols) for the filament of width $w_a$. The solid and dashed blue lines correspond to the hybrid law using the parameters $(v_{max},\theta_0)$ and $(v'_{max},\theta'_0)$, respectively. The gray lines correspond to the Cox--Voinov law.}
\label{fig:xf-h0-thex_17_num}
\end{figure}

The solid lines in Fig.~\ref{fig:xf-h0-thex_17_num} show the time dependence of the tip position, $x_f$, tip contact angle, $\theta_x$, and maximum head thickness, $h_0$, as given by the numerical simulations using the values of the constants $\theta_0$ and $v_{max}$ from Eqs.~(\ref{eq:theta0}) and (\ref{eq:vmax}), respectively. Regarding Fig.~\ref{fig:xf-h0-thex_17_num}(a), the simulations yield smaller values of both the slope of $x_f(t)$ for early times and the maximum tip position for late times respect to the experimental data. Note also, that the numerical values of $\theta_x$ and $h_0$ lie below the experimental points (see Figs.~\ref{fig:xf-h0-thex_17_num}(b) and (c)).

Clearly, the degree of agreement between simulations and experiments is highly dependent on the parameters in Eq.~(\ref{eq:theta_fit_num}). The weakest part of this fitting process is the inclusion of a single point to account for an important part of the dewetting region of this constitutive relation, namely $v_{max}$, which is obtained by performing a quite different experiment than that used to generate the rest of the data (see Fig.~\ref{fig:theta_vcl}). Since there is an uncertainty in the value of $v_{max}$ due to the experimental error (of the order of $7\%$), we explore the possibility that its inaccuracy can be a source of these differences. Moreover, although there is consensus on the hypothesis that the contact angle is zero at $v_{max}$~\cite{petrov_cs85}, one cannot guarantee that this is actually the case. If a different contact angle corresponds to this velocity, then the resulting $\theta(v_{cl})$ curve allows both higher angles and dewetting velocities. The simulations show that considering a $7\%$ increase in $v_{max}$ strongly improves the comparison of the slope of $x_f(t)$ for early times, while the saturation value is not practically not affected. On the other hand, there is an additional uncertainty on how the value of $\theta_0$ is determined,  since the only requirement is that it falls inside the hysteresis range, namely $(\theta_a,\theta_d)$. In this case, the numerical results show that the saturation value of $x_f(t)$ favorably compares with the experiments if $\theta_0$ is increased a few degrees with respect to the value given in Eq.~(\ref{eq:theta0}). 

Thus, in order to better approximate the experimental data, we perform a simulation by slightly modifying both parameters to the values $v'_{max}=1.07 v_{max}$ and $\theta'=50.57^\circ$, and using consistent modified values of the coefficients in Eq.~(\ref{eq:param}) to fit the experimental data in Fig.~\ref{fig:theta_vcl}: $\Gamma'= 95.4553$, $\ell'= 0.0008302 a_c=1.24 \times 10^{-4}$~cm, $v'_0=6.2121 \times 10^{-7}$~cm/s. The results of the simulation are shown by the dashed lines in Fig.~\ref{fig:xf-h0-thex_17_num}. We observe a clear improvement of the comparison for all three quantities shown in that figure, in particular for the evolution of the contact angle at the tip, $\theta_x$.

\begin{figure}
\centering
\subfigure[$t=0$~s]
{\includegraphics[width=0.4\linewidth]{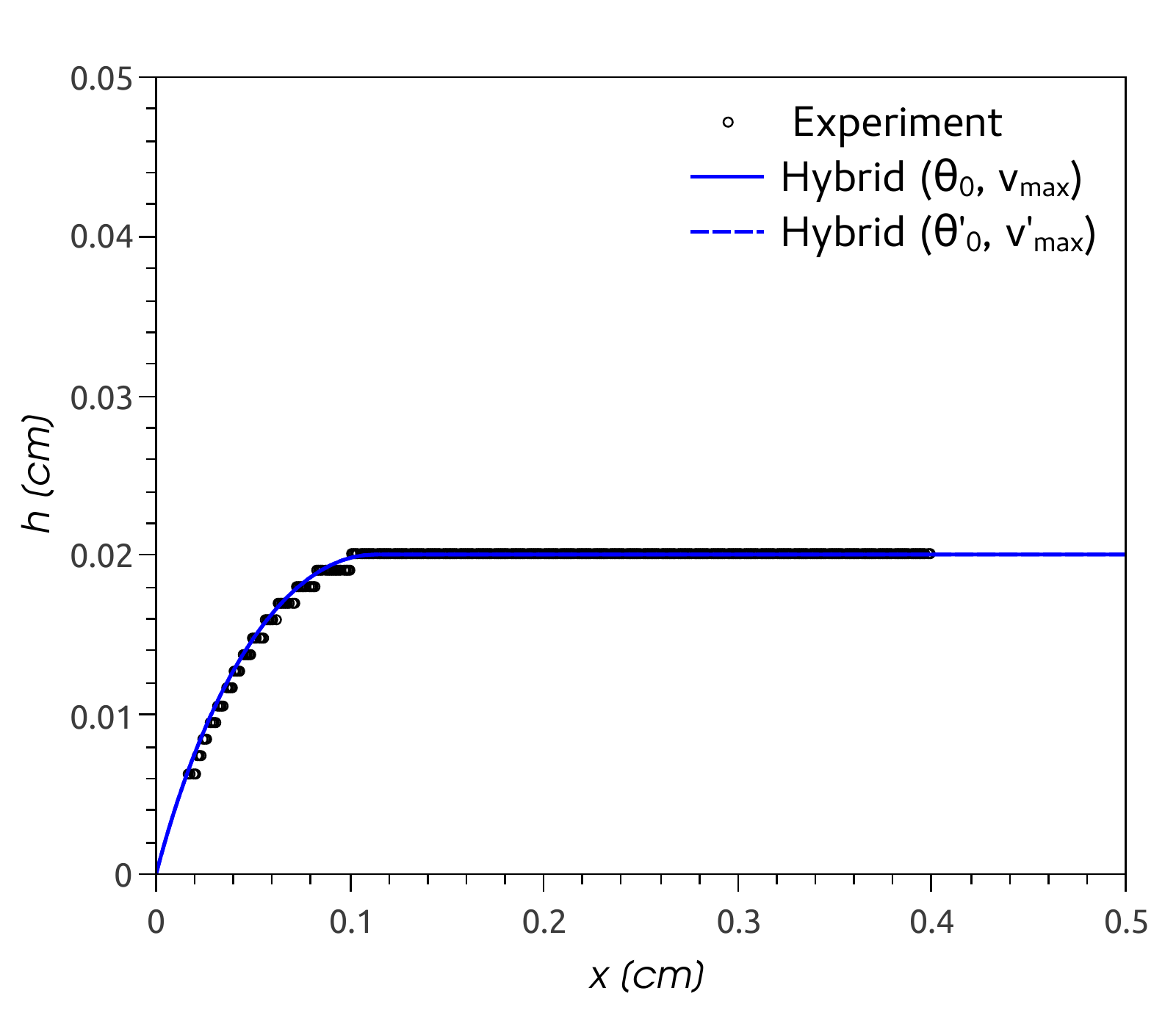}}
\subfigure[$t=30$~s]
{\includegraphics[width=0.4\linewidth]{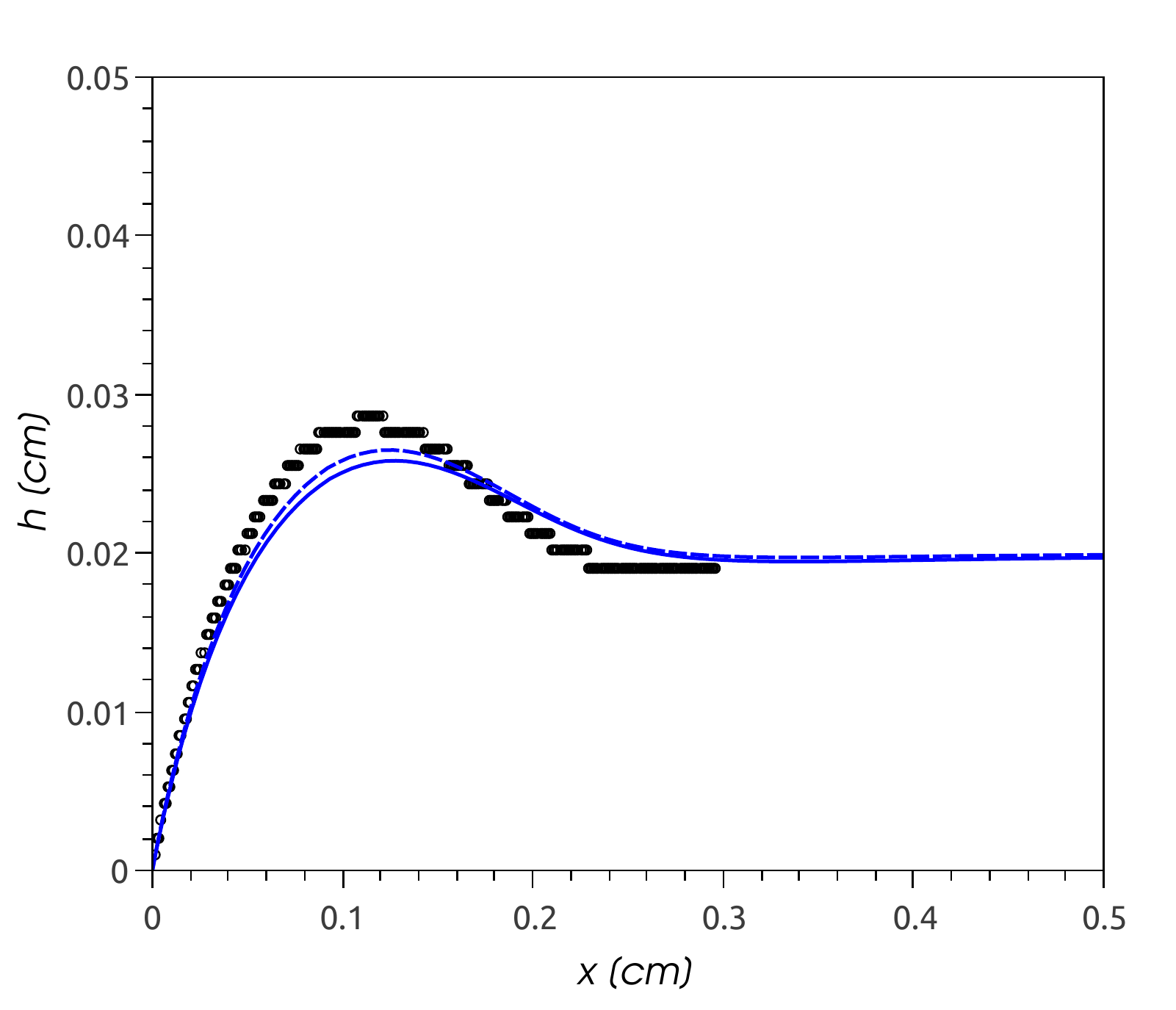}}\\
\subfigure[$t=130$~s]
{\includegraphics[width=0.4\linewidth]{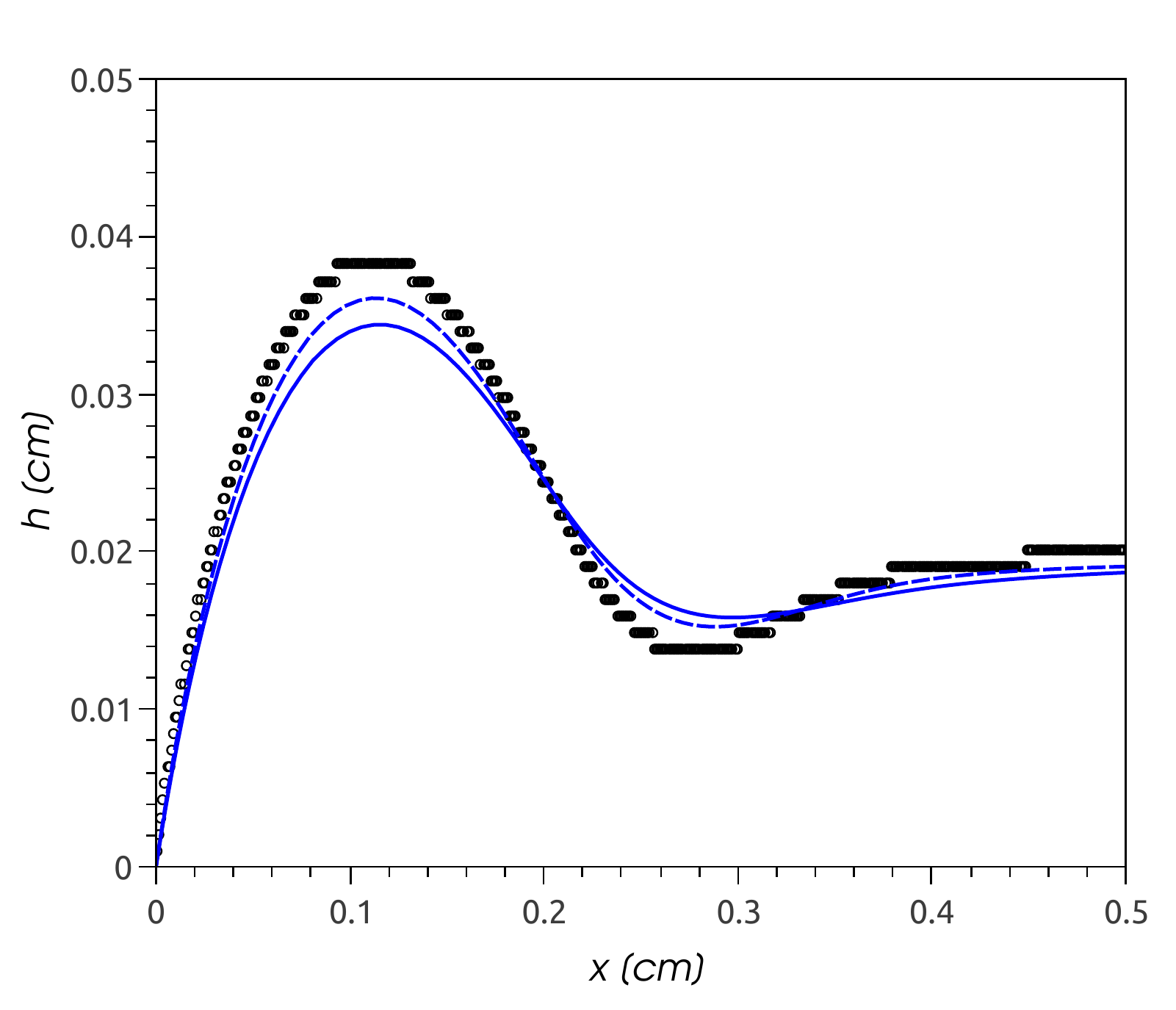}}
\subfigure[$t=165$~s]
{\includegraphics[width=0.4\linewidth]{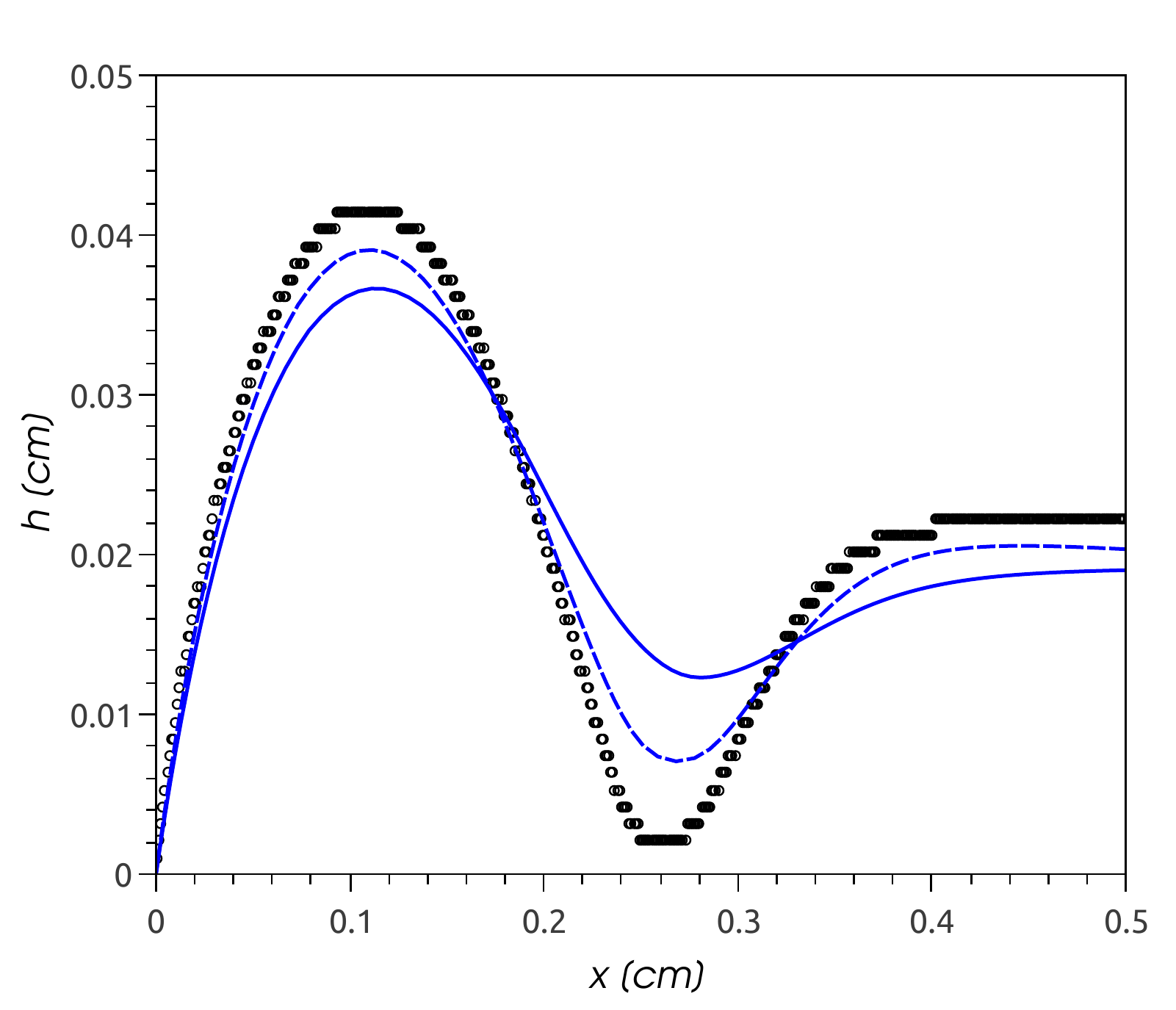}}
\caption{{\it Side view}. Thickness profiles for the filament of width $w_a$ at different times. The black dots correspond to the experimental, while the solid and dashed blue lines correspond to numerical profiles using the parameters $(v_{max},\theta_0)$ and $(v'_{max},\theta'_0)$, respectively.}
\label{fig:hx_comp}
\end{figure}

A similar tendency to approach the experimental data is evident in the thickness profiles when using the alternative values $v'_{max}$ and $\theta'_0$ (see Fig.~\ref{fig:hx_comp}). In order to focus our attention on the shape of the profiles, both the experimental and numerical ones have been shifted so that the position of the tip is always at $x=0$. Although, there are some departures both near the maximum and the neck, the latter are the most significant ones.

\begin{figure}
\centering
\subfigure[]
{\includegraphics[width=0.45\linewidth]{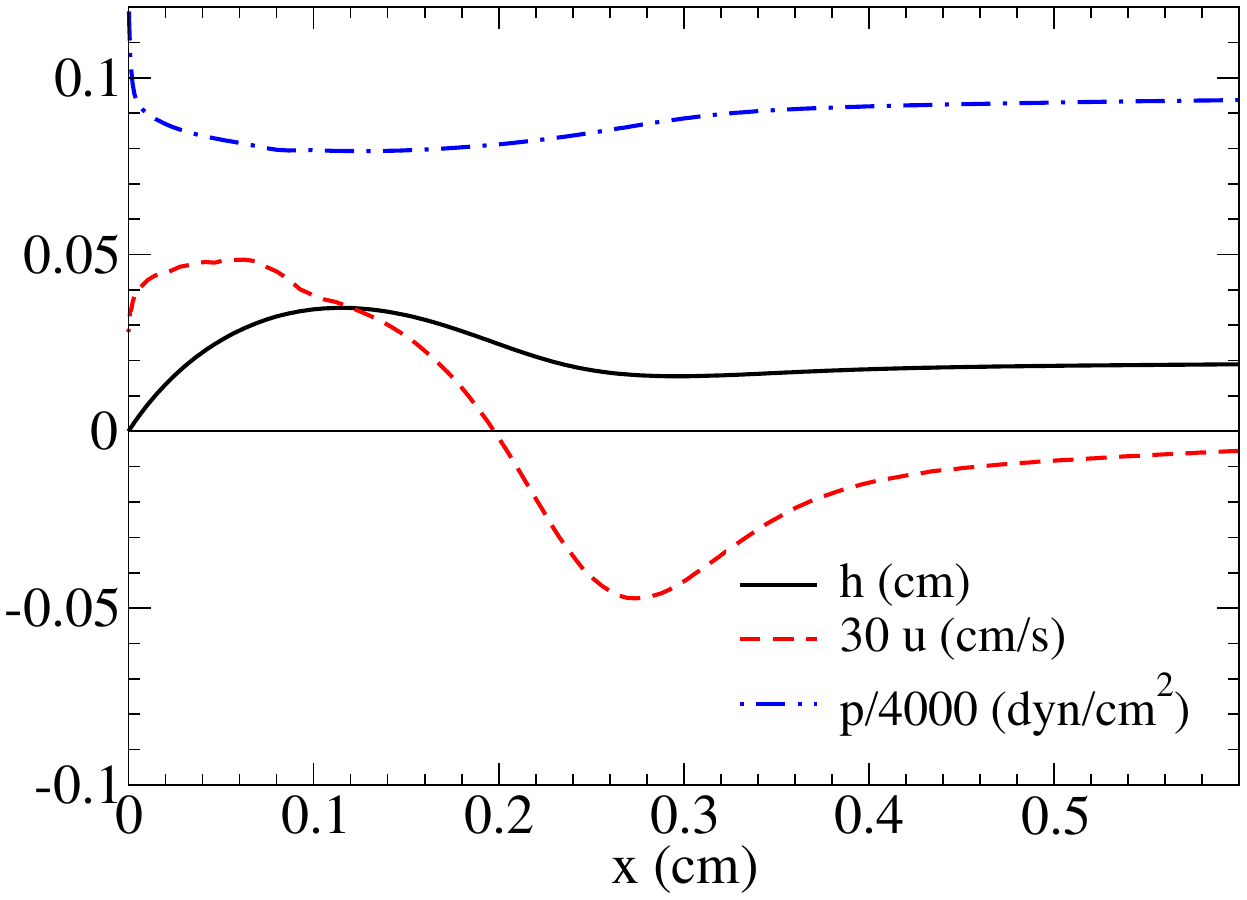}}
\subfigure[]
{\includegraphics[width=0.45\linewidth]{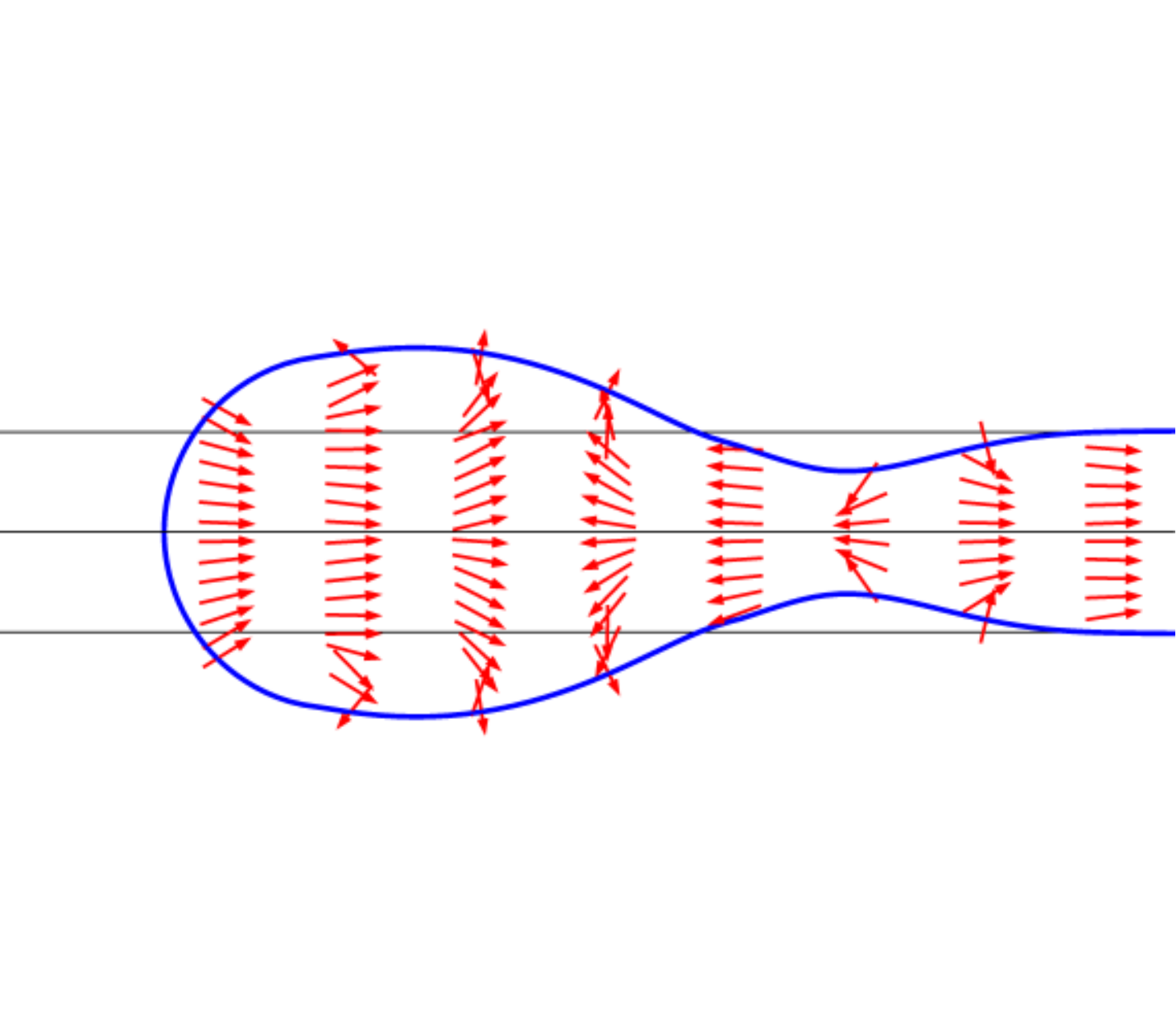}}
\caption{Numerical results for the head and neck regions for width $w_a$. (a) Dimensional profiles of thickness, $h$, axial velocity, $u$, and pressure, $p$, at the free surface in a vertical plane containing the filament axis ($y=0$) at $t=900 t_c$ . (b) Velocity field at a horizontal plane near the substrate at $t=1300 t_c$ (i.e. close to the breakup time).}
\label{fig:hvp_num}
\end{figure}

Since the differences between experiments and numerics are mainly quantitative and reasonable small, the latter are still very useful to learn about the hydrodynamics of the dewetting process at the filament end. In fact, the numerical results in Fig.~\ref{fig:hvp_num}(a) show the behaviour of the velocity and pressure fields in the head and neck regions for times close to breakup at the intersection of the free surface with the plane $y=0$. At the tip, the pressure is relatively high and induces the dewetting motion ($u<0$). Inside the head, the pressure is practically uniform and increases towards the neck, where it reaches a maximum value. This peak drives a flow out of the neck region towards the head ($u<0$) and the filament ($u>0$). 

Fig.~\ref{fig:hvp_num}(b) shows the flow field in a horizontal plane close to the substrate. This vectorial field velocity complements the picture already described and indicates that the change of behaviour from receding to advancing occurs at different values of $x$ whether one observes the problem at the top of the free surface or near the contact line.

\begin{figure}[hbt]
\centering
\subfigure[]
{\includegraphics[width=0.45\linewidth]{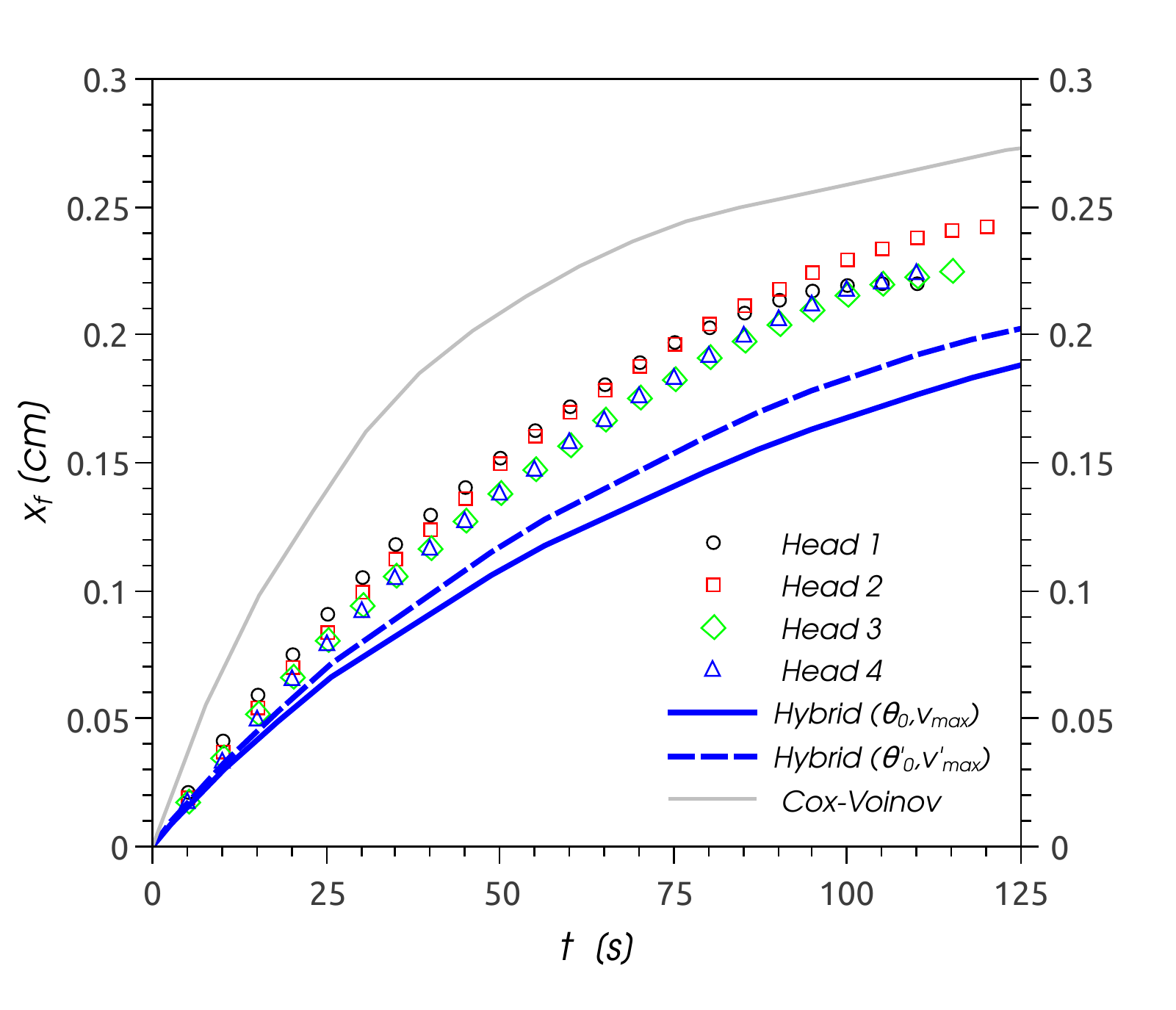}}
\subfigure[]
{\includegraphics[width=0.45\linewidth]{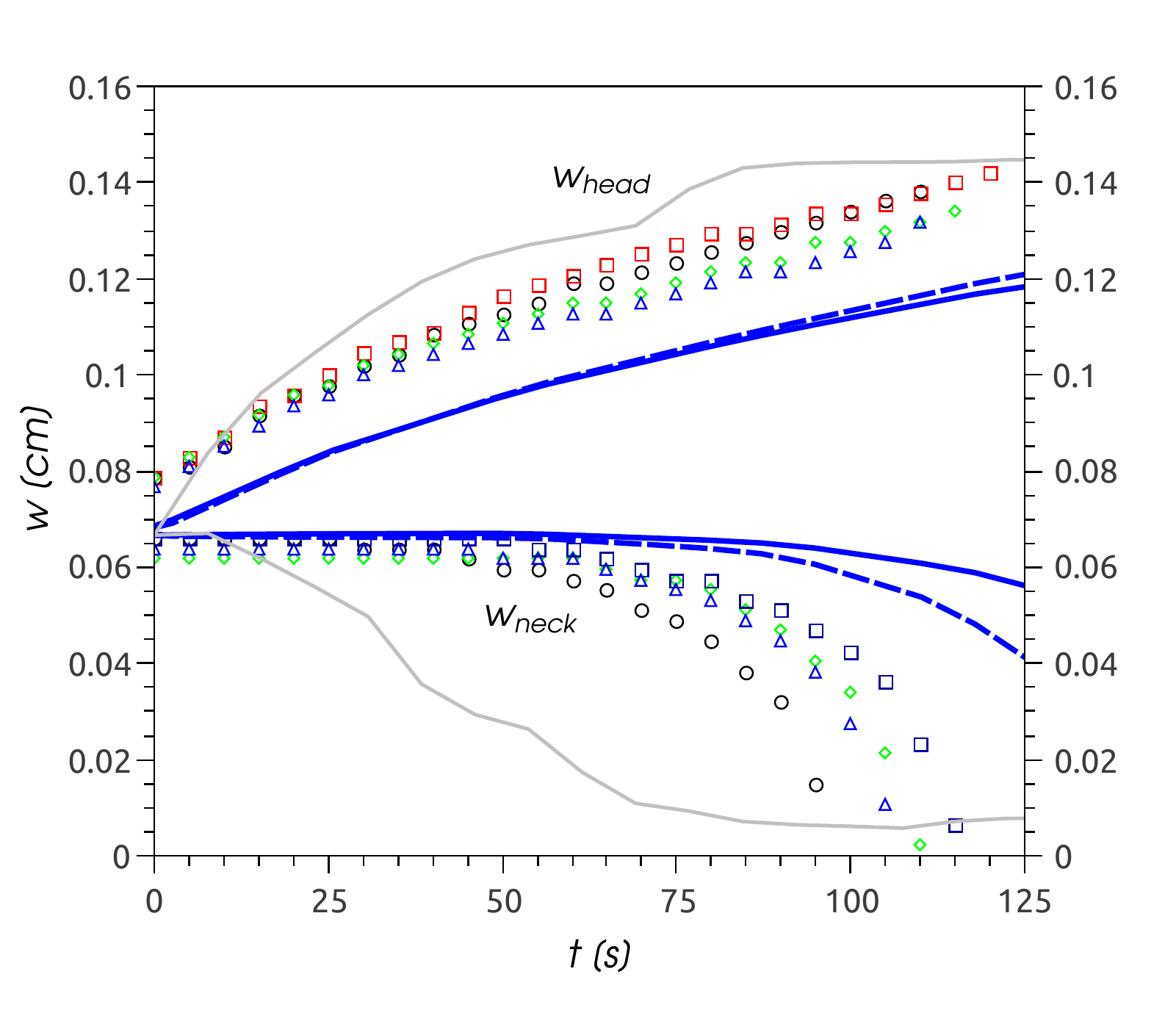}}
\caption{{\it Top view}. Comparison of the numerical solution with the experimental data in Fig.~\ref{fig:xf-w_20} (symbols) for the filament of width $w_b$. The solid and dashed blue lines correspond to the hybrid law using the parameters $(v_{max},\theta_0)$ and $(v'_{max},\theta'_0)$, respectively. The gray lines correspond to the Cox--Voinov law. (a) Time evolution of the tip position, $x_f$. (b) Head and neck width, $w_{head}$ and $w_{neck}$, respectively.}
\label{fig:xf-w_20_num}
\end{figure}

In order to describe the evolution of the shape of the head, we resort now to an experiment observed from the top, see Fig.~\ref{fig:side_top}(b), which allows us to register the evolution of the whole contact line. In Fig.~\ref{fig:xf-w_20_num}, we compare the tip position, $x_f(t)$, the head width, $w_{head}$, and the neck width, $w_{neck}$, with the experimental data. The comparison of $x_f(t)$ yield similar conclusions with respect to the effects of using the modified values $(v'_{max},\theta'_0)$ to those of the side view case in Fig.~\ref{fig:xf-h0-thex_17_num}(a). However, unlike the previous case we observe here that the slope of $x_f(t)$ for early times is a bit smaller than the experimental results. Also, Fig.~\ref{fig:xf-w_20_num}(b) shows a greater departure of the widths of both head and neck from the experiments, as well as very slight improvements when using the primed parameters. We believe that all these differences can be attributed to the fact that the ellipsoidal shape assumed at $t=0$ for the filament end after breakup might not be the best choice to describe the longitudinal thickness profile. Unfortunately we are not able to observe simultaneously both the thickness profile and the shape of the footprint to completely elucidate this issue.

\begin{figure}
\centering
\subfigure[$t=0$~s]
{\includegraphics[width=0.4\linewidth]{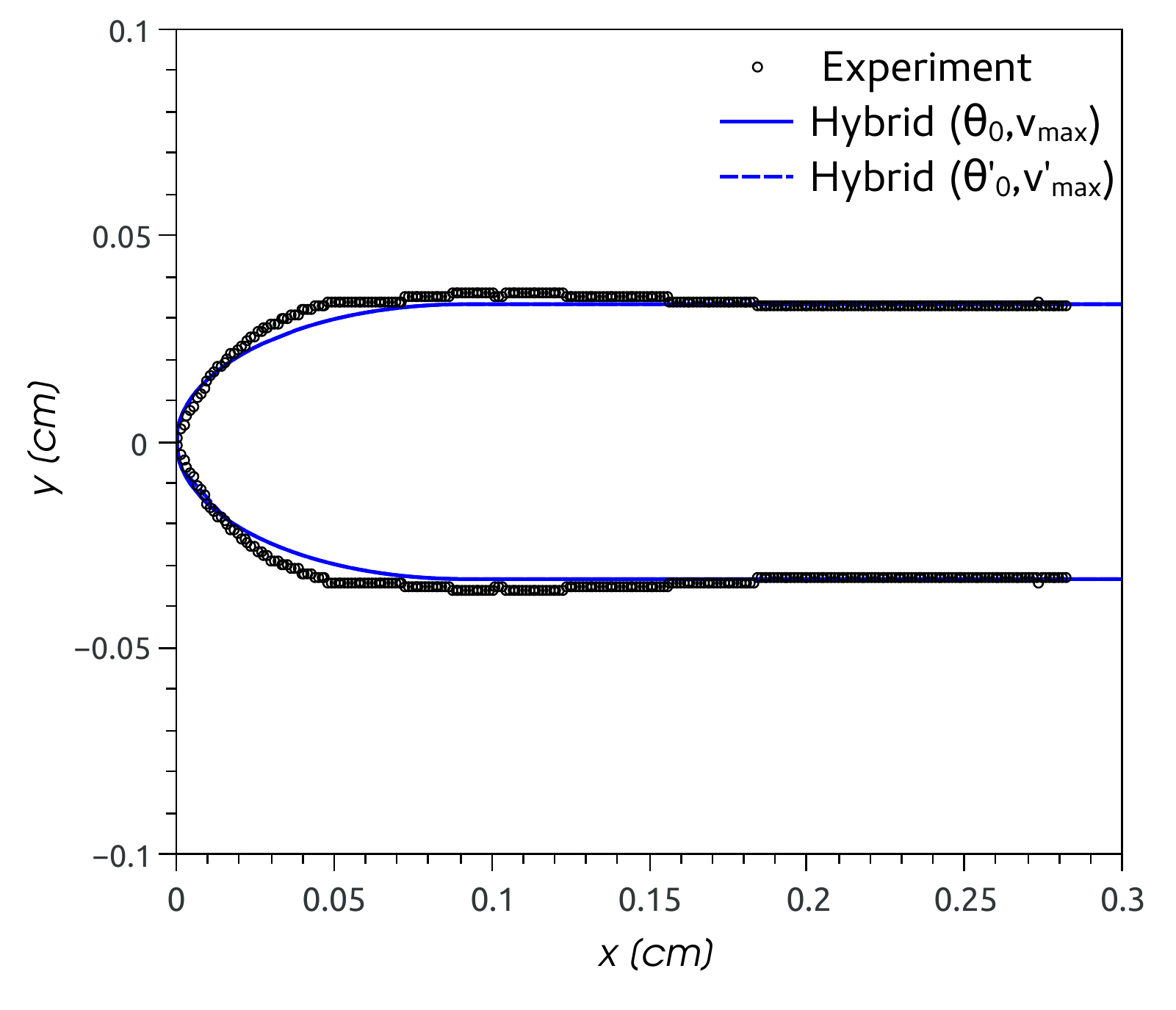}}
\subfigure[$t=45$~s]
{\includegraphics[width=0.4\linewidth]{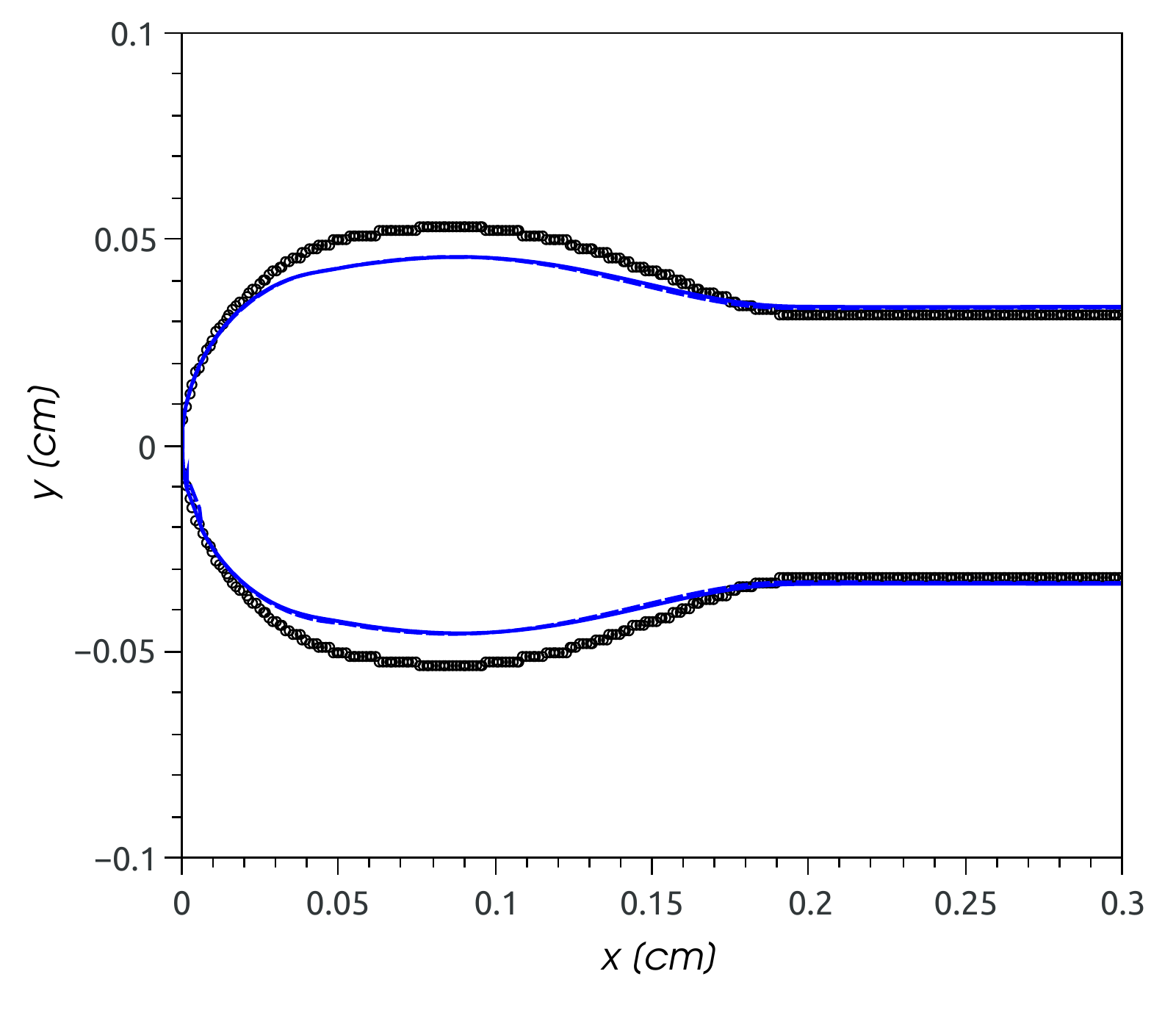}}\\
\subfigure[$t=95$~s]
{\includegraphics[width=0.4\linewidth]{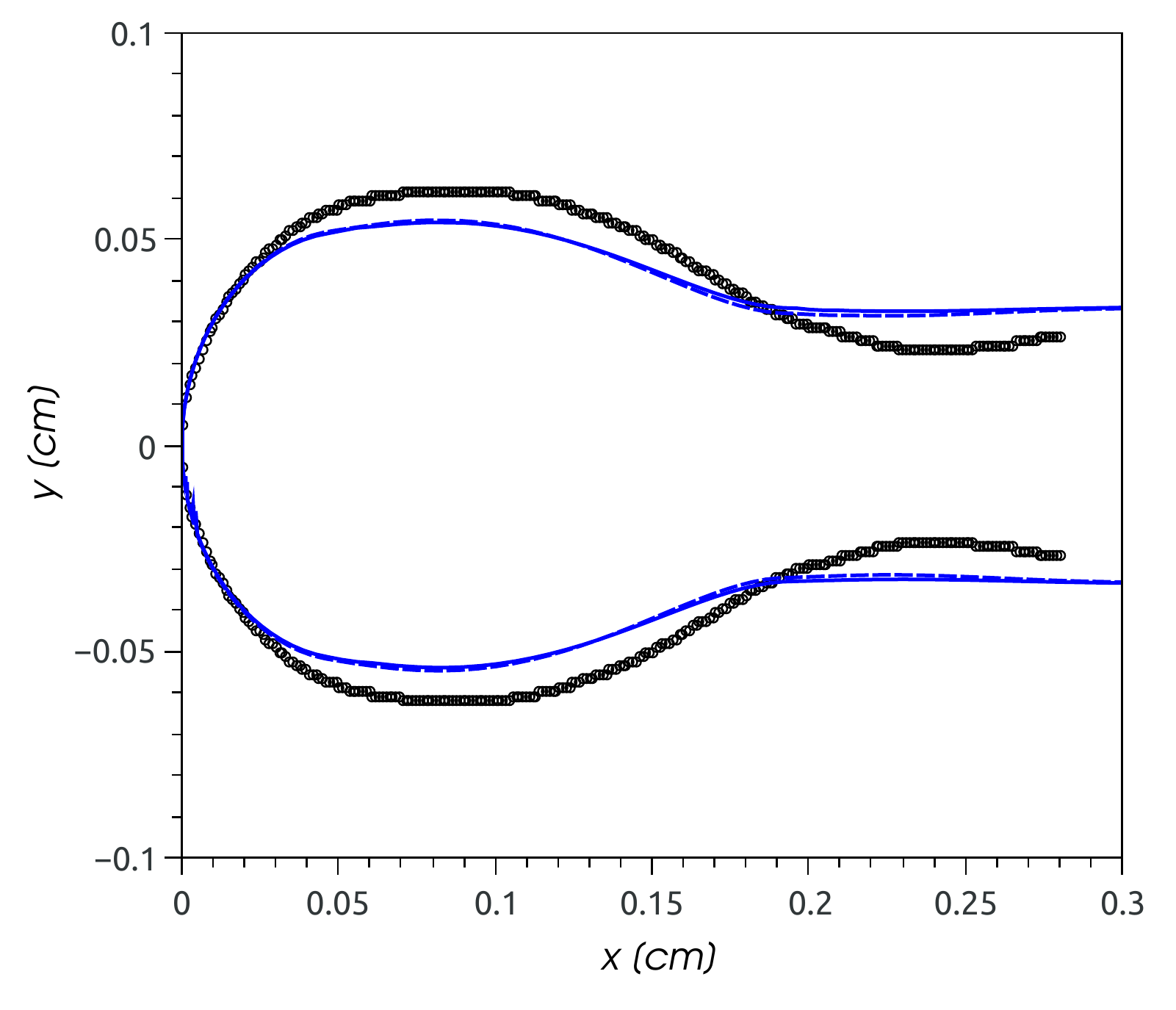}}
\subfigure[$t=115$~s]
{\includegraphics[width=0.4\linewidth]{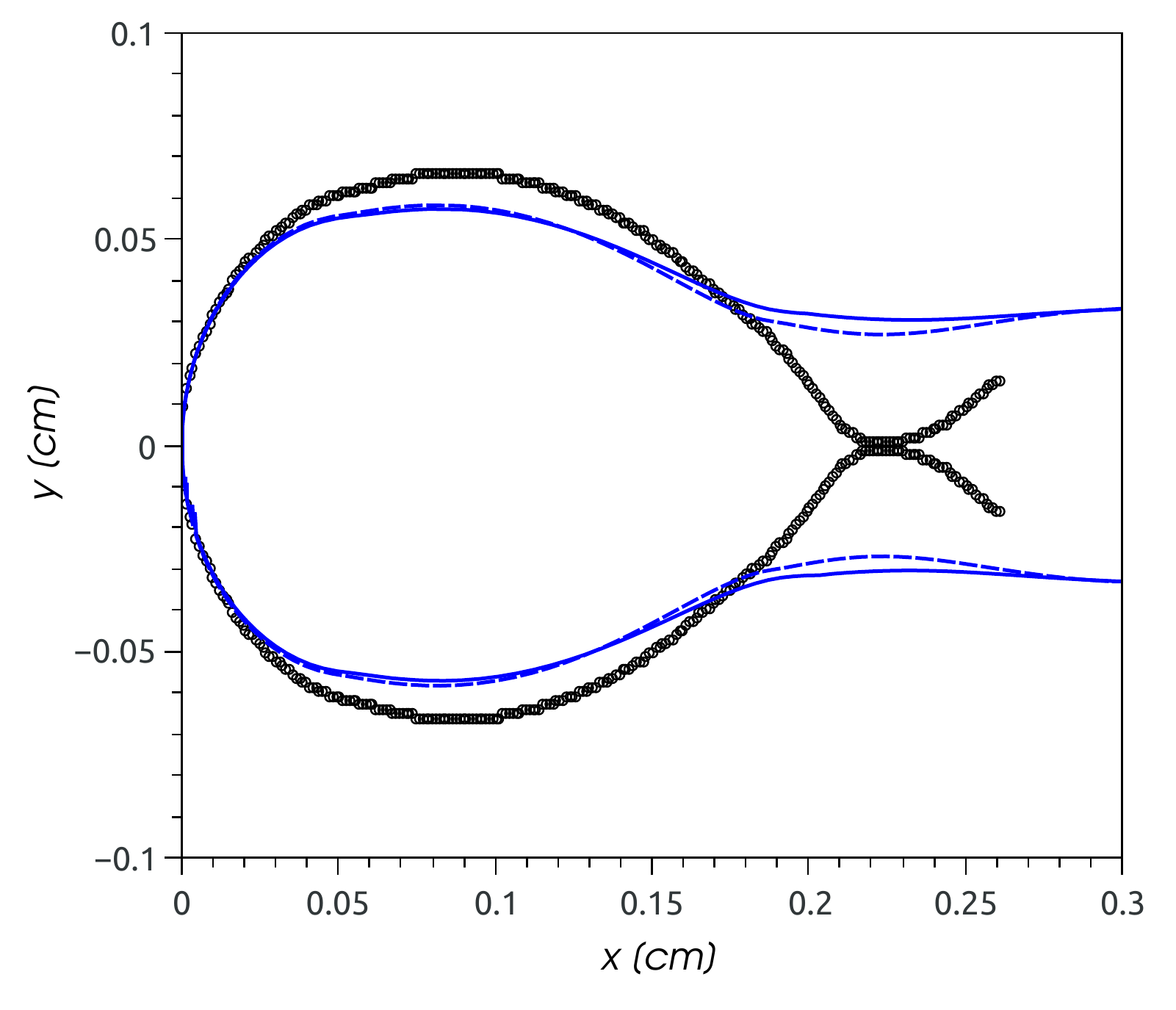}}
\caption{{\it Top view}. Contours (contact line shapes) for the filament of width $w_b$ at different times. The black dots correspond to the experimental data, while the solid and dashed blue lines correspond to numerical contours using the parameters $(v_{max},\theta_0)$ and $(v'_{max},\theta'_0)$, respectively.}
\label{fig:foot_comp}
\end{figure}

In Fig.~\ref{fig:foot_comp}, we compare the contours of the footprints at different times by using $(v_{max},\theta_0)$ (solid lines) and $(v'_{max},\theta'_0)$ (dashed lines), as we have done for the thickness profiles in Fig.~\ref{fig:hx_comp}. Here, we also focus our attention on the shape of these footprints, so that both the experimental and numerical contours have been shifted so that the position of the tip is always at $x=0$. Unlike the comparison made for the thickness profiles, the modified parameters yield here a slight improvement only in the prediction of the neck shape, while the shape of the rest of the contact line is barely affected.

\begin{figure}
\centering
\subfigure[$t=0-1100 \,t_c$]
{\includegraphics[width=0.45\linewidth]{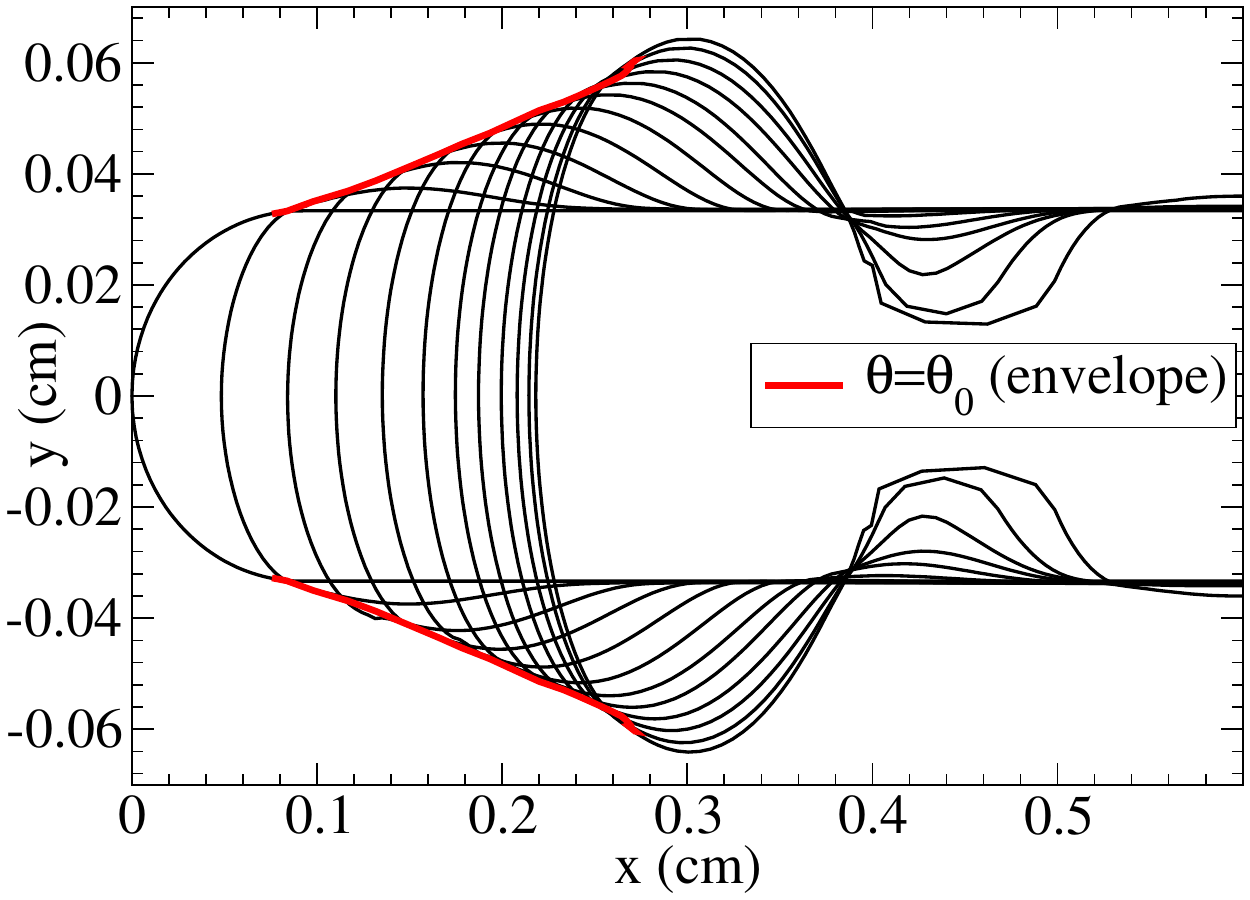}}
\subfigure[$t=900\,t_c$]
{\includegraphics[width=0.45\linewidth]{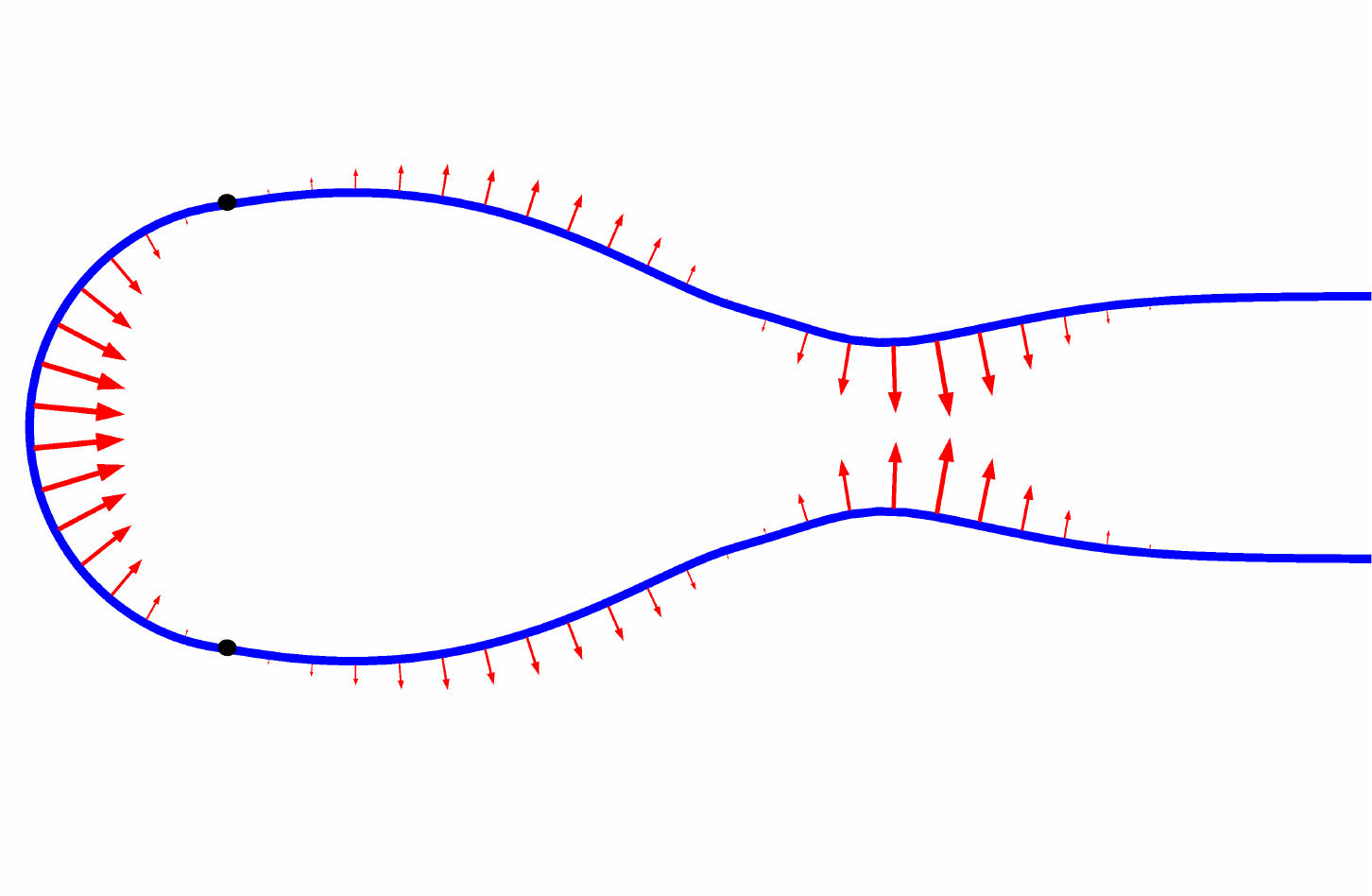}}
\caption{(a) Time evolution of the contact line (thin black lines), and trajectories of the contact line point where $\theta=\theta_0$ (red thick lines). (b) Contact lines velocity field showing the point where the front motion changes direction from advancing to receding.}
\label{fig:env_sim}
\end{figure}

Interestingly, the simulations also show the existence of an angular sector that embraces the evolution of the head contact line, similarly to what is observed in the experiments (see Fig.~\ref{fig:the_vcl-fil}(b)). This feature is shown in Fig.~\ref{fig:env_sim}(a). In fact, the red thick lines in that figure correspond to the envelope of the parametric family of curves $(x(s,t),y(s,t))$ representing the contact line in the simulations. Here, the envelope is calculated by numerically solving the condition:
\begin{equation}
 \frac{\partial x}{\partial t}\frac{\partial y}{\partial s}-\frac{\partial x}{\partial s}\frac{\partial y}{\partial t}=0.
 \label{eq:wronk}
\end{equation}
Note this curve can be approximated by a straight line, as the one observed in the experiments and whose slope turned out to be independent of $w$. Additional simulations show that this is actually the case, though the constant $\alpha$ is $7.69^\circ$ here. In accordance with Fig.~\ref{fig:foot_comp}, this angle is not modified by using the set of parameters $(v'_{max},\theta'_0)$. The difference with the experimental value of $\alpha$ is consistent with the departures of $x_f$ and $w_{head}$ mentioned above.

On the other hand, we calculate the curve joining the contact line points where the local contact angle is equal to $\theta_0$, i.e. where the contact line velocity is zero. The interesting fact is that this other curve turns out to be identical to the above envelope. Thus, we conclude that the existence of the envelope is related to the points of the contact line which do not have normal velocity, but only tangential one. This effect is illustrated by the black points in Fig.~\ref{fig:env_sim}(b), where the contact line velocity field is shown by vectors. Therefore, these points divide the portion of the contact line that is receding (dewetting)  from that portion  which is advancing (wetting) in the normal direction to the contact line. This result allows to understand the nature of the envelope that embraces the evolving footprint of the head and provides the physical underlying reason of its existence.

\section{Summary and conclusions}
\label{sec:conclu}
In this paper we are concerned with the dynamics of the axial dewetting of a liquid filament laying on a horizontal substrate under partial wetting conditions. We observe that the retraction motion as well as the consequent transverse spreading strongly depend on the features of the contact angle hysteresis. For this reason, we study in detail the hysteresis cycles both under static and dynamic conditions. In the first case, we are able to determine the hysteresis range as well as the critical angles at which the drop contact line position must change in order to obtain a new static shape as its volume is varied (see Section~\ref{sec:st_wet}). For the dynamic case, we obtain the constitutive relationship between the contact line velocity and the dynamic contact angle, and include this relation within a hybrid model that takes into account both a hydrodynamic approach and a molecular kinetic description (see Section~\ref{sec:dyn_wet}). These results play a fundamental role in understanding the dynamics of the retracting filament, which is the main goal of this work.

The measurements of different characteristics of the evolving filament, such as tip position, contact angle at the tip, head thickness, head width, and neck width as well as contact line shape and vertical thickness profile, allow to perform a complete description of the whole evolution, so that the proposed wettability model can be thoroughly tested. In order to understand the main features of this flow, we develop a heuristic model to account for the time evolution of these parameters. The main outcome of this model is a remarkably good prediction of the retraction motion of the tip and the contact angle there. However, it is unable to properly describe the time evolution of the width of the head, since it does not consider the properties of the contact angle hysteresis in the transverse direction.

Therefore, we also perform a comparison of the experimental data with the results of the numerical solution of the full Navier--Stokes equations, which include the contact line velocity versus the contact angle as given by the hybrid constitutive relationship resulting from Eqs.~(\ref{eq:theta_hyd}) and (\ref{eq:theta_bl}). The simulations are able to properly describe the flow when using this hybrid law with the coefficients as measured in the characterization of the hysteresis cycle. It should be pointed out that the use of this function $\theta(v_{cl})$ is an essential ingredient to adequately account for the experimental data. Naturally, one could be tempted to use a more usual and simpler relation as given by Eq.~(\ref{eq:theta_hyd}) with $\theta_m=\theta_0$, known as Cox--Voinov model. However, the simulations show that using this law instead of the hybrid one leads to results that do not compare well with the experiments (see gray lines in Figs.~\ref{fig:xf-h0-thex_17_num} and \ref{fig:xf-w_20_num}). This fact strongly emphasizes the need to properly model the hysteretic effects when describing the dynamics of a receding filament, which simultaneously involves wetting and dewetting motions of the contact line.

As discussed in Section~\ref{sec:num}, the main differences with the simulations observed in some experimentally measured quantities for the retracting filament end can be attributed to an incomplete knowledge of its shape just after the breakup. One of the most interesting facts revealed by the simulations is the physical property of the contact line points which belong to the envelope that encloses the footprints at different times. It is shown that these points have no normal velocity of the contact line, and correspond to the characteristic contact angle of the hysteresis curve, $\theta_0$. This envelope is also observed in the experiments as two straight lines, in agreement with the simulations.

Our analysis has focused on the retraction mechanism of the filament prior to its breakup. However, the final stages of this process can be affected by the rupture dynamics, whose understanding certainly merits future work.

\acknowledgments
The authors acknowledge support from Consejo Nacional de Investigaciones Cient\'{\i}ficas y T\'ecnicas (CONICET, Argentina) with grant PIP 844/2012 and Agencia Nacional de Promoci\'on Cient\'{\i}fica y Tecnol\'ogica (ANPCyT, Argentina) with grant PICT 931/2012. We also thank Dr. Mathieu Sellier (Canterbury Univ., NZ) for his contributions in the numerical simulations.

\end{document}